\documentclass[fleqn,usenatbib]{mnras}

\usepackage[T1]{fontenc}

\DeclareRobustCommand{\VAN}[3]{#2}
\let\VANthebibliography\thebibliography
\def\thebibliography{\DeclareRobustCommand{\VAN}[3]{##3}\VANthebibliography}

\usepackage{graphicx}	
\usepackage{amsmath}	
\usepackage{amssymb}	
\newcommand{\angstrom}{\mbox{\normalfont\AA}}
\usepackage{multirow}
\usepackage{newtxtext,newtxmath}

\title[Kinematics of the BLR in AGNs]{Kinematics of the H$\alpha$ and H$\beta$ broad line region in an SDSS sample of type 1 AGNs}

\author[N. Raki\'c]{
N. Raki\'c,$^{1,2}$\thanks{E-mail: nemanja.rakic@pmf.unibl.org}
\\
$^{1}$Physics Department, Faculty of Natural Sciences and Mathematics, University of Banjaluka, Mladena Stojanovi\'ca 2, 78000 Banjaluka, RS, Bosnia and Herzegovina\\
$^{2}$Department of Astronomy, Faculty of Mathematics, University of Belgrade, Studentski trg 16, 11000 Belgrade, Serbia\\
}

\date{Accepted XXX. Received YYY; in original form ZZZ}

\pubyear{2022}

\begin{document}
\label{firstpage}
\pagerange{\pageref{firstpage}--\pageref{lastpage}}
\maketitle

\begin{abstract}
Here we investigate the kinematics of the part of the broad line region (BLR) in active galactic nuclei (AGNs) emitting  H$\beta$ and H$\alpha$ emission lines. We explore the widths and asymmetries of the broad H$\beta$ and H$\alpha$ emission lines in a sample of high quality (i.e. high signal to noise ratio) spectra of type 1 AGN taken from the Data Release 16 of the Sloan Digital Sky Survey, in order to explore possible deviation from the gravitationally bound motion. To find only the broad component of H$\beta$ and H$\alpha$ we use the FANTASY (Fully Automated pythoN Tool for AGN Spectra analYsis) code for the multi-component modeling of the AGN spectra and for careful extraction of the broad emission line parameters. 
We show  that based on the broad line profiles widths and asymmetries, the BLR gas emitting H$\beta$ and H$\alpha$ lines follows similar kinematics, and seems to be virialized in our sample of type 1 AGN.
\end{abstract}

\begin{keywords}
line: profiles - galaxies: active - quasars: emission lines
 
\end{keywords}



\section{Introduction}
Determination of the mass of the supermassive black holes (SMBH) residing in the center of most if not all galaxies that have a bulge component, is an important factor for understanding the galaxy evolution and the interplay of the SMBH and its host galaxy  \citep[see e.g., a review ][and references therein]{2013ARA&A..51..511K}. However, measuring the SMBH mass is still a complex task based on either resolving the region of the SMBH gravitational influence, for direct dynamical methods, or using various scaling relations for indirect mass estimates \citep[see e.g.,][]{2014SSRv..183..253P}. All of which are strongly dependant on the kinematics of the region which properties are being measured.

In case of galaxies hosting an active galactic nuclei (AGN), which could be traced to much higher redshift due their higher luminosity, the SMBH mass can be measured using broad emission lines,  observed in so-called type 1 AGN \citep[for a recent review see e.g.,][]{pop20}. One well-known method using the width of the broad emission-lines, assumes the that line-emitting gas (so-called broad-line region - BLR), surrounding the SMBH is virialized, i.e., the kinematics of the BLR gas is governed by the gravity of the SMBH \citep[see e.g.,][etc.]{1999ApJ...521L..95P,2000ARA&A..38..521S,2015ARA&A..53..365N}. However, there are indications that the BLR kinematics is more complex, due to possible presence of several emitting components \citep[e.g.,][]{2004A&A...423..909P}, or radial motions, such as inflows or outflows \citep[e.g.,][]{2009NewAR..53..140G,2020A&A...644A.175V}.

Usually, the virilization of the BLR is considered a priori and is widely used as an assumption when estimating the SMBH mass in AGNs from spectral parameters, i.e. the broad line width and the BLR radius. Probing if the BLR gas kinematics is gravitationally bound to the SMBH, i.e. if the virialization holds and what would be the observational signatures to support this, is important for understanding when the BLR properties, namely the emission line widths, could be used to measure the SMBH mass \citep[e.g.,][]{2012NewAR..56...49M,jon,2019MNRAS.484.3180P}. This has great implications for the wide usage of scaling relations involving the emission line widths and fluxes to measure the SMBH mass from a single-epoch spectrum \citep[][]{2002MNRAS.337..109M, 2020ApJ...903..112D}. In case of low-luminosity near-by AGN, empirical correlations for the so-called virial mass based on the broad H$\alpha$ emission line are widely used \citep[][]{2005ApJ...630..122G, 2011ApJ...739...28X}. For most AGNs, the width of H$\beta$ line was mostly used as a virial broadening estimator of the SMBH mass \citep[][]{2001MNRAS.327..199M,2006ApJ...641..689V}, e.g. it was used in the catalogue of Sloan Digital Sky Survey (SDSS) type 1 AGN \citep[][]{2011ApJS..194...45S}. For more distant AGN, one should use UV lines, such as C IV \citep[][]{2011ApJ...742...93A} and Mg II \citep[][]{2016MNRAS.460..187M}. Therefore, it is important to find different ways to test if the line-emitting gas is virialized. 

\cite{2019MNRAS.484.3180P} investigated the kinematics of the H$\beta$ and Mg II line emitting regions, and showed that the H$\beta$ and Mg II core component seems to be virialized, concluding that Mg II could be used for the SMBH mass estimates if the Mg II core component is dominant. \cite{2022arXiv220507034M} studied the virialization in UV lines, namely Al III $\lambda $1860 doublet and the C III] $\lambda$1909 line, and showed that they could be used as virial broadening estimators, similar to H$\beta$. Here we aim to investigate the kinematics of the H$\alpha$ and H$\beta$ broad-line emitting-region by careful extracting the pure broad line profiles, and measuring the line widths and asymmetries. We aim to perform simultaneous multicomponent spectral fitting in order to deblend the broad H$\alpha$ and H$\beta$ line from the  satellite narrow lines and Fe II lines. 

For this study, we use a sample of high quality (i.e. high signal to noise ratio) optical spectra collected for 946 low-redshift type 1 AGNs from the Sloan Digital Sky Survey (SDSS) Data Release 16 \citep{dr16}. We divide AGNs into two population A and B, based on a limit on the H$\beta$ line full width at half maximum (FWHM) $\sim$ 4000 km/s, as around this width the clear change in the shape of the H$\beta$ line profile is seen \citep[][]{2002ApJ...566L..71S}. Population A and B objects are discussed to have significantly different physical properties \citep[see e.g.,][and reference therein]{2000ARA&A..38..521S,2022AN....34310082M}. These populations occupy different areas of the FWHM H$\beta$ - $R_{\rm Fe II}$ space, which is known as the quasar main sequence primarily driven by Eddington ratio convolved with the line of sight orientation\citep[][]{2000ARA&A..38..521S,2014Natur.513..210S,2022AN....34310082M}. The parameter $R_{\rm Fe II}$ is a measure of the optical Fe II emission, defined as the ratio of equivalent widths of Fe II emission in the range 4435–4685 \AA \, and H$\beta$ broad line. 
 In summary, Population B sources are showing broader emission line widths and weaker Fe II emission ($R_{\rm Fe II} \leq 0.5$), most likely having low Eddington ratio and being viewed at higher inclination angles. On the other hand, Population A sources are mostly associated with higher Eddington ratio, and show narrower broad emission lines and stronger Fe II emission ($R_{\rm Fe II} \geq 0.5$). These are believed to be viewed more face-on \citep[see for a recent review][]{2022AN....34310082M}. We use the separation to population A and B object following the argument that the change in the line profile occurs after 4000 km s$^{-1}$ \citep[][]{2018rnls.confE...2M}, as well as to have equally populated sub-samples.

The paper is organized as follows: section 2 describes the data sample and the performed analysis, section 3 gives the theoretical justification of our approach, section 3 presents and discuss the obtained results , and finally, section 4 outlines the conclusions.
For calculating the luminosity through the luminosity distance, we used the cosmological parameters ${\rm H}_0$ = 70 km s$^{-1}$ Mpc$^{-1}$ , $\Omega_m$ = 0.30, and $\Omega_\Lambda$ = 0.70.

\section{Data \& Methods}\label{sec2}

\subsection{Sample of type 1 AGNs}

In order to retrieve a large sample of type 1 AGN having high-quality spectra, i.e. with high signal-to-noise (S/N) ratio, we explored the SDSS Data Release 16 \citep[DR16,][]{dr16}. DR16 is the fourth data release of the fourth phase of the survey (SDSS-IV), including the complete data set of optical single-fibre spectroscopy of the SDSS through February 2019 \citep[see][for details on SDSS spectrographs and spectral resolution]{2013AJ....146...32S}. These data represent the culmination of SDSS 20 year mission in collecting optical spectra to map the 3D structure of the Universe \footnote{\href{www.sdss.org/dr16/spectro/}{www.sdss.org/dr16/spectro/}}. 
We used Structured Query Language (SQL) to query the SDSS DR16 SkyServer search tools\footnote{\href{www.skyserver.sdss.org/dr16/en/tools/search/sql.aspx}{www.skyserver.sdss.org/dr16/en/tools/search/sql.aspx}} data table ‘specobjALL’, choosing spectra that follow these three simple criteria:
\begin{enumerate}
    \item redshift $z<0.4$ with $z_{\text{warning}}=0$ to ensure that both H$\beta$ and H$\alpha$ are included, i.e. spectra cover the wavelength region at least up to 7000 \AA;
    \item signal to noise ratio in g-band higher than 30, to ensure high-quality spectra around H$\beta$ emission lines;
    \item objects classified as "QSO".
\end{enumerate}
The SQL query returned 960 objects. After visual inspection of preliminary spectral fittings (see Section 2.3), we excluded 14 objects with poor fitting results, typically due to bad pixels near emission lines of interest, or strong cosmic ray presence.
This left 946 objects in the sample which was further studied in more details. Further in the analysis, we divide the total sample into two sub-samples based on the broad H$\beta$ FWHM to: i) population B with FWHM(H$\beta$) > 4000 km/s, resulting with 526 objects, and ii) population A with FWHM(H$\beta$) < 4000 km/s giving 420 objects. Furthermore, we discuss as a subset of population A object, so-called extreme population A, with strong iron emission and R$_{\rm FeII}$ > 1 \citep[][]{2022AN....34310082M}, which consist of 82 objects.

\begin{figure}
\centerline{\includegraphics[width=1\columnwidth]{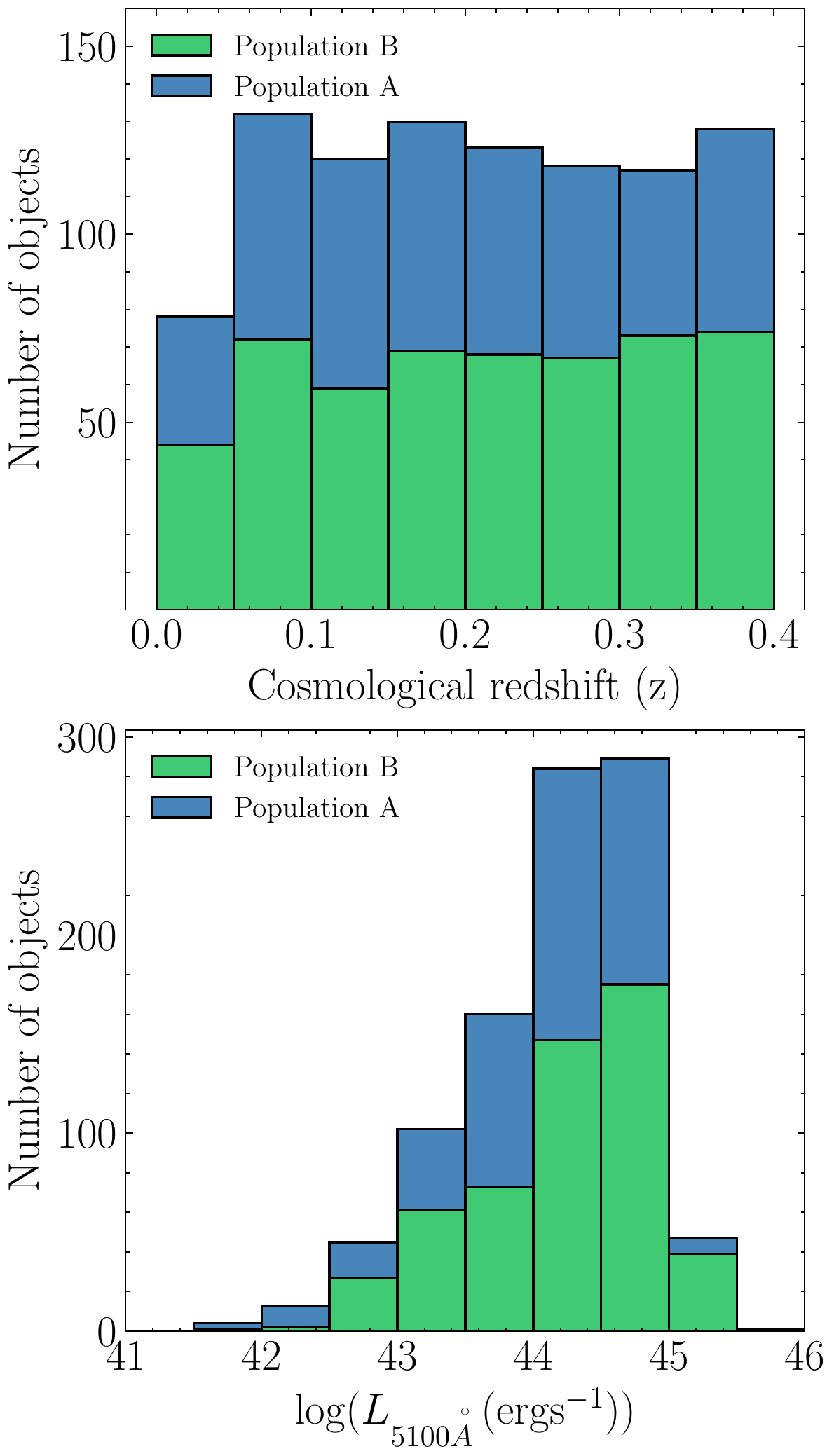}}
\caption{Histograms of the cosmological redshift (upper)  and continuum luminosity at 5100 $\angstrom$ (bottom) showed as stacked bars of population A (blue) and population B (green) objects.\label{fig1}}
\end{figure}

\subsection{Properties of the selected sample} 

We briefly discuss the properties of our sample of 946 type 1 AGN. Fig. \ref{fig1} give histograms of the cosmological redshift $z$ (upper panel) and continuum luminosity $L_{5100\angstrom}$ (lower panel) for the total sample, shown as stacked bars of population A (blue) and population B (green) objects. Our total sample is uniformly sampled across the selected range of redshift, as well as both subsets. Most of the objects in the total sample (95\%) have the continuum luminosity log($L_{5100}$ erg s$^{-1}$) in the range [42.65-45.11]. The distribution is having the median log$L_{5100\angstrom}$=44.30 and is asymmetric toward higher-luminosity AGN, as is expected for the sample of SDSS type 1 AGNs \citep[see][and their Figure 9]{2019ApJS..243...21L}. 

We tested if different selection of S/N ratio influence the sample properties. Putting less strict limit for S/N>20, the redshift distribution remains the same, and there is no significant influence on the distribution of luminosities, whereas the sample size increase roughly 3 times. However, since the main aim of this work is to carefully extract the broad H$\alpha$ and H$\beta$ emission lines and measure their properties, lowering the limit to S/N will increase the presence of noisy spectra, which will increase the scatter of measured emission line-parameters.

Therefore, the final selected sample consists of the AGNs with high-quality spectra, in which we can analyse the emission-line shapes very precisely.

\begin{figure}
    \centering
    \includegraphics[width=\columnwidth]{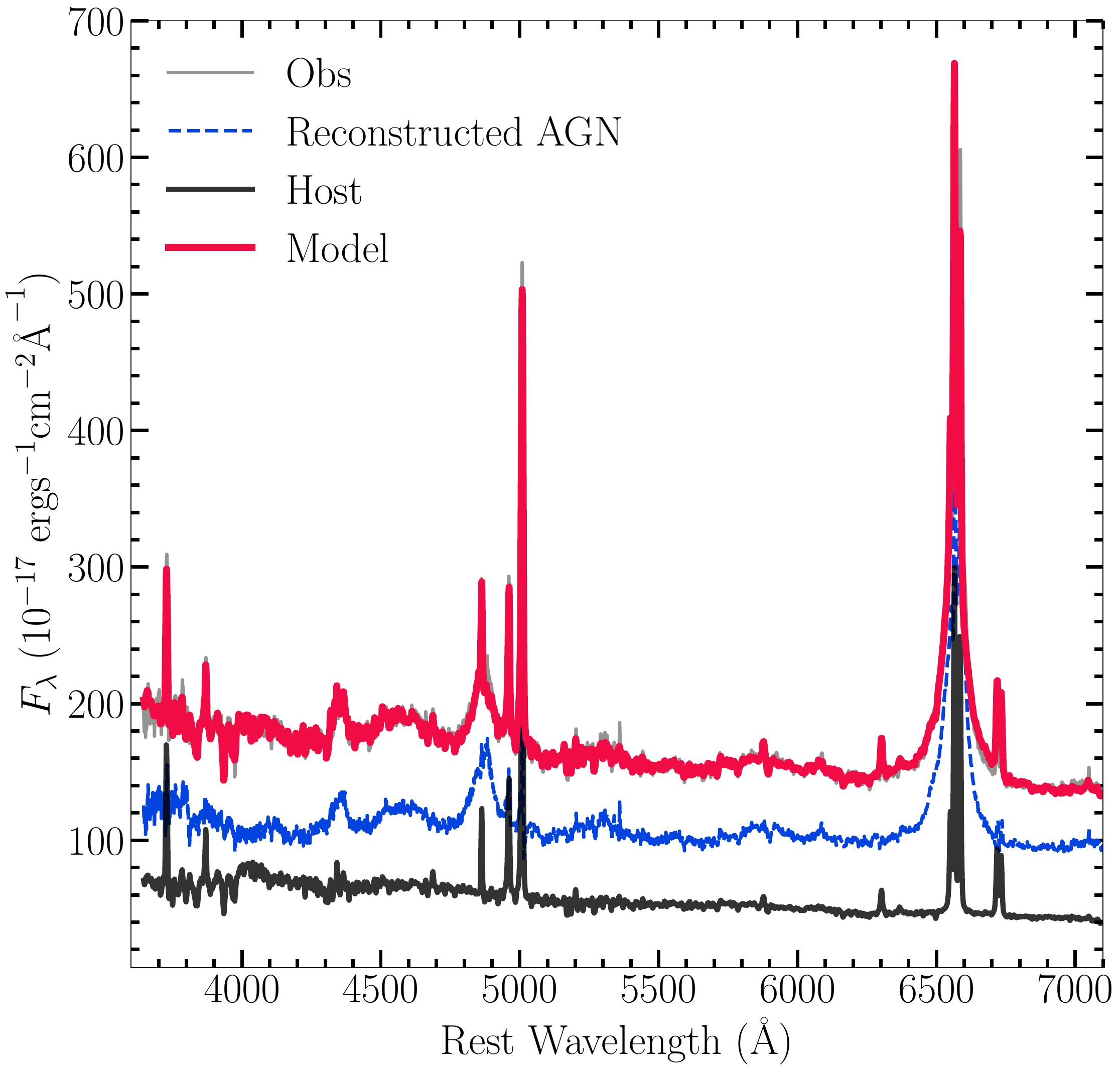}
    \caption{Example of the PCA host reconstruction using in case of the object SDSS J101912.56+635802.6. Observed flux is denoted with gray color, the reconstructed model with red, host galaxy flux is given with green, and pure reconstructed AGN (observed flux - host galaxy flux) with blue.}
    \label{fig:host}
\end{figure}

\subsection{Spectral fittings and analysis} 

For the AGN spectral analysis, we used python-based code for multi-component spectral fitting - FANTASY (Fully Automated pythoN Tool for AGN Spectra analYsis)\footnote{This is an open-source code, soon to be available on GitHub}, first used in \cite{fant}. 

The first step is the preparation of spectra for the fittings. We corrected the spectra for Galactic extinction using dust map data from \cite{redden}, as well as for the cosmological redshift using the redshift provided in the SDSS database. Intending to get pure AGN spectra, we firstly have removed host galaxy contamination by applying the PCA method explained in \cite{berk}. According to \cite{berk} majority of spectra can be reconstructed using the linear combination of 10 QSO eigenspectra from \cite{yipb} and 5 galaxy eigenspectra from \cite{yipa}.
We remove host galaxy contribution by subtracting the reconstructed host galaxy from the observed spectrum. Example of host reconstruction is given in the Fig. \ref{fig:host}.

\begin{figure}
    \centering
    \includegraphics[width=\columnwidth]{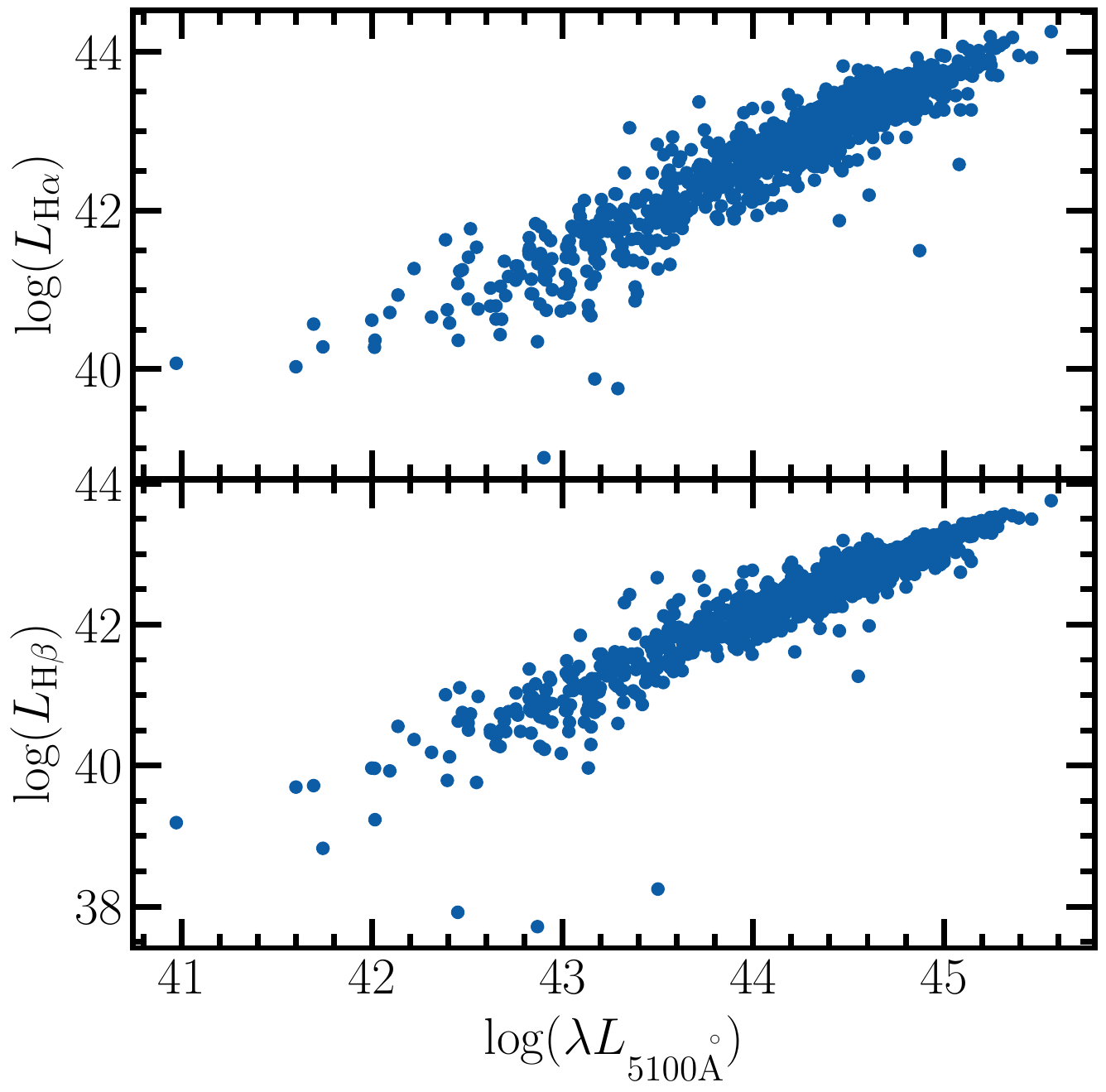}
    \caption{Luminosity of the extracted broad H$\alpha$ (upper panel) and H$\beta$ (bottom panel) line vs. AGN continuum luminosity $L_{5100\angstrom}$, in erg/s.}
    \label{fig:lum}
\end{figure}

\begin{figure*}
    \centering
    \includegraphics[width=0.97\textwidth]{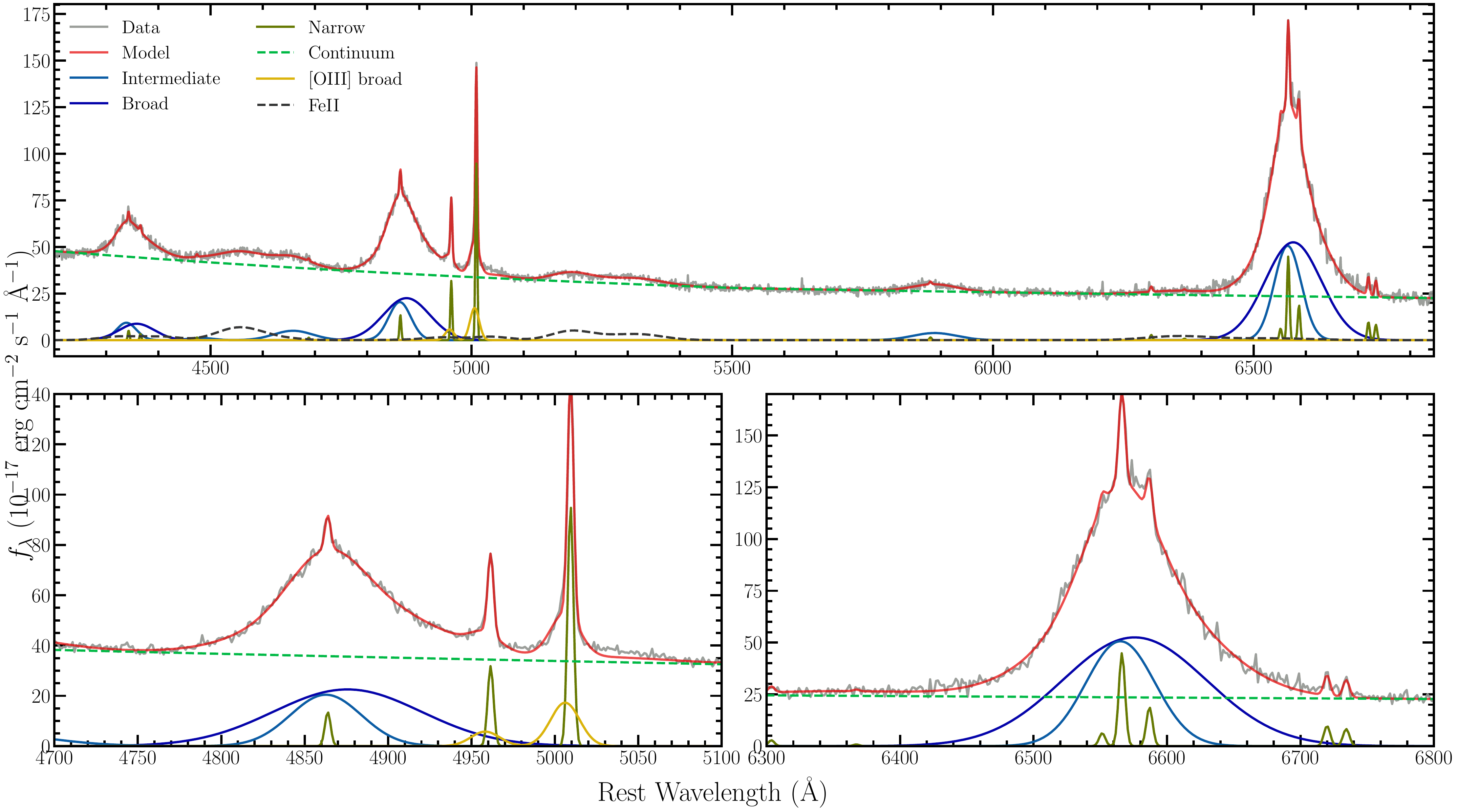}
    \caption{Multi-component simultaneous fitting of the $\lambda\lambda$4200-7000 \AA\AA\, wavelength region for the object SDSS J074910.59+284214.5, which is an example of Population B object. Below the observed (gray) and modeled (red) spectrum, the broken power-law continuum (green dashed) and all Gaussian components are shown (blue, dark-blue - broad components of H$\alpha$ and
      H$\beta$; green - narrow lines; yellow - [O III] broad component). Black dashed line indicate the Fe II model. Below two panels show the zoom-in of the H$\beta$ and H$\alpha$ line region.}
    \label{fig:fit}
\end{figure*}

\begin{figure*}
    \centering
    \includegraphics[width=\textwidth]{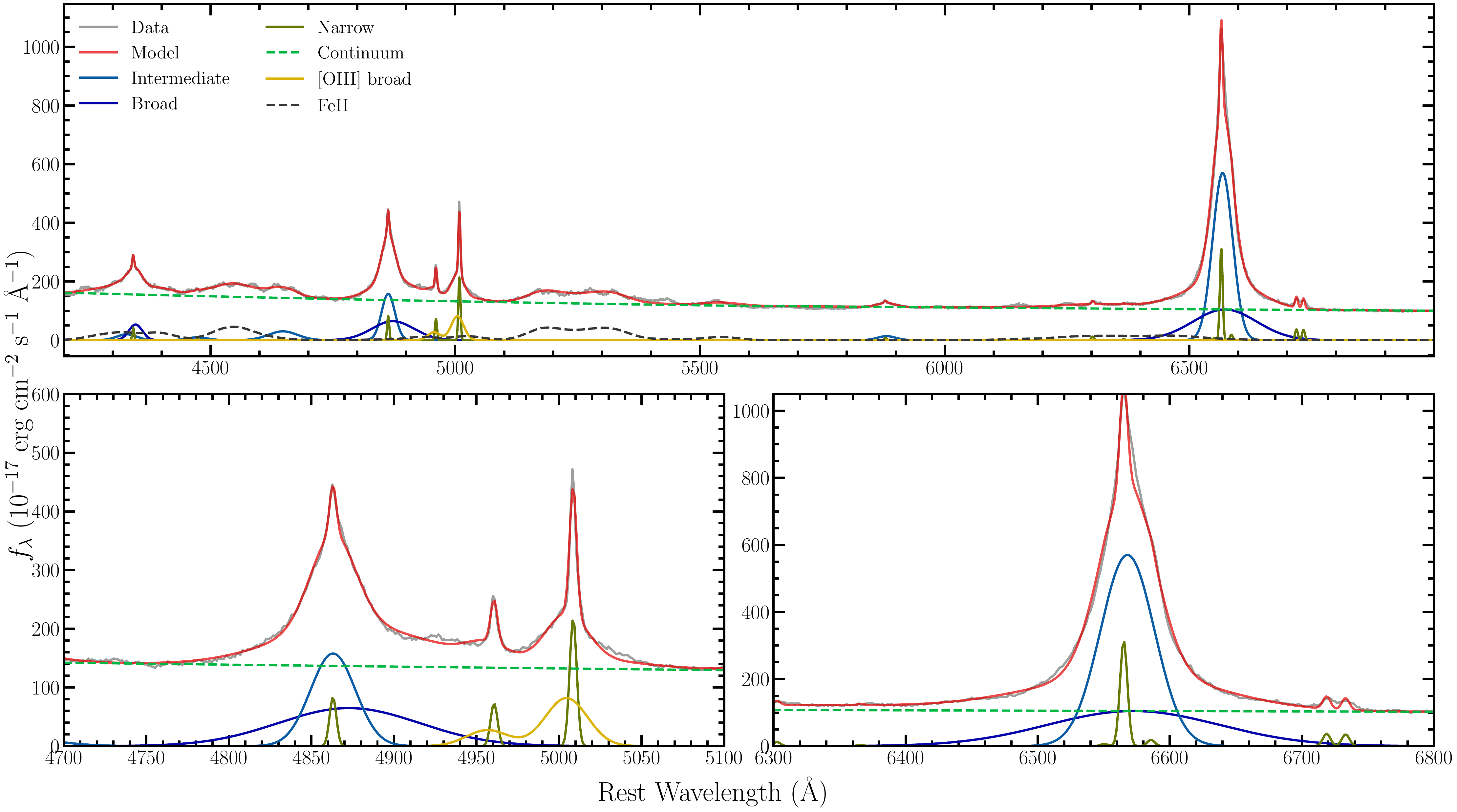}
    \caption{The same as in Figure \ref{fig:fit} but for the object SDSS J135550.20+204614.5, which is an example of the Population A object.}
    \label{fig:fit_nls}
\end{figure*}

Further, we applied simultaneous multi-component spectral fitting to remove narrow and satellite lines to extract the broad component of both H$\beta$ and H$\alpha$ lines\footnote{Spectral fittings were performed using the SUPERAST computer cluster of the University of Belgrade - Faculty of Mathematics, Department of Astronomy \citep[][]{2022PASRB..22..231K}.}. We fitted all spectra in the rest wavelength range $\sim$4200-7000 \AA\AA\ using the same model. The used model consisted from: 
\begin{enumerate}
    \item broken power law continuum;
    \item hydrogen H$\alpha$, H$\beta$ and H$\gamma$ lines with three Gaussian components (two broad and one narrow);
    \item helium lines He I 4471$\mathrm{\mathring{A}}$, He I 5877$\mathrm{\mathring{A}}$, and He II 4686$\mathrm{\mathring{A}}$;
    \item narrow emission lines, all fixed to have the same shifts and widths as [O III] 5007$\mathrm{\mathring{A}}$: [O III] 4363$\mathrm{\mathring{A}}$, [O III] 4959, 5007$\mathrm{\mathring{A}}\mathrm{\mathring{A}}$, [N II] 6548, 6583$\mathrm{\mathring{A}}\mathrm{\mathring{A}}$, [S II] 6716, 6731$\mathrm{\mathring{A}}\mathrm{\mathring{A}}$; the ratio of [O III] 4959,5007$\mathrm{\mathring{A}}\mathrm{\mathring{A}}$ and [N II] 6548,6583$\mathrm{\mathring{A}}\mathrm{\mathring{A}}$ doublets were fixed to 3 \citep[][]{2007MNRAS.374.1181D,2022arXiv220410036D}; the list of used narrow lines contains also nebular line [O I] 6300$\mathrm{\mathring{A}}$ and [O I] 6364$\mathrm{\mathring{A}}$, but these were seldom identified;
\item the broad component of the [O III] doublet \citep[][]{2022A&A...659A.130K}, 
\item optical Fe II model based on the atomic data of the transitions
\citep[see for details][]{kov10,2012ApJS..202...10S,kov15}. 
\end{enumerate}
The total sample was fitted automatically with this model. We have visually inspected all the results of the fittings. To illustrate the performance of the automated spectral fittings, we plot in Figure \ref{fig:lum}, how the luminosity of the extracted broad line H$\alpha$ and H$\beta$ fluxes behaves with the continuum luminosity $L_{5100}$ \AA. As expected by the photoionization theory explaining the origin of the broad emission lines \citep[][]{2006agna.book.....O,2013peag.book.....N}, as well as shown by observations \citep[e.g.,][]{2017FrASS...4...12I,2020ApJ...903..112D}, the broad emission line fluxes are strongly correlated with the AGN continuum flux. This supports that the fitting results are reasonable. Examples of the fits for a population B object is given in Fig. \ref{fig:fit}, and for Population A in Fig. \ref{fig:fit_nls}. The goodness of the best fitting results were evaluated using the $\chi^2$ parameter. 

\begin{figure}
\centering
\includegraphics[width=0.95\columnwidth]{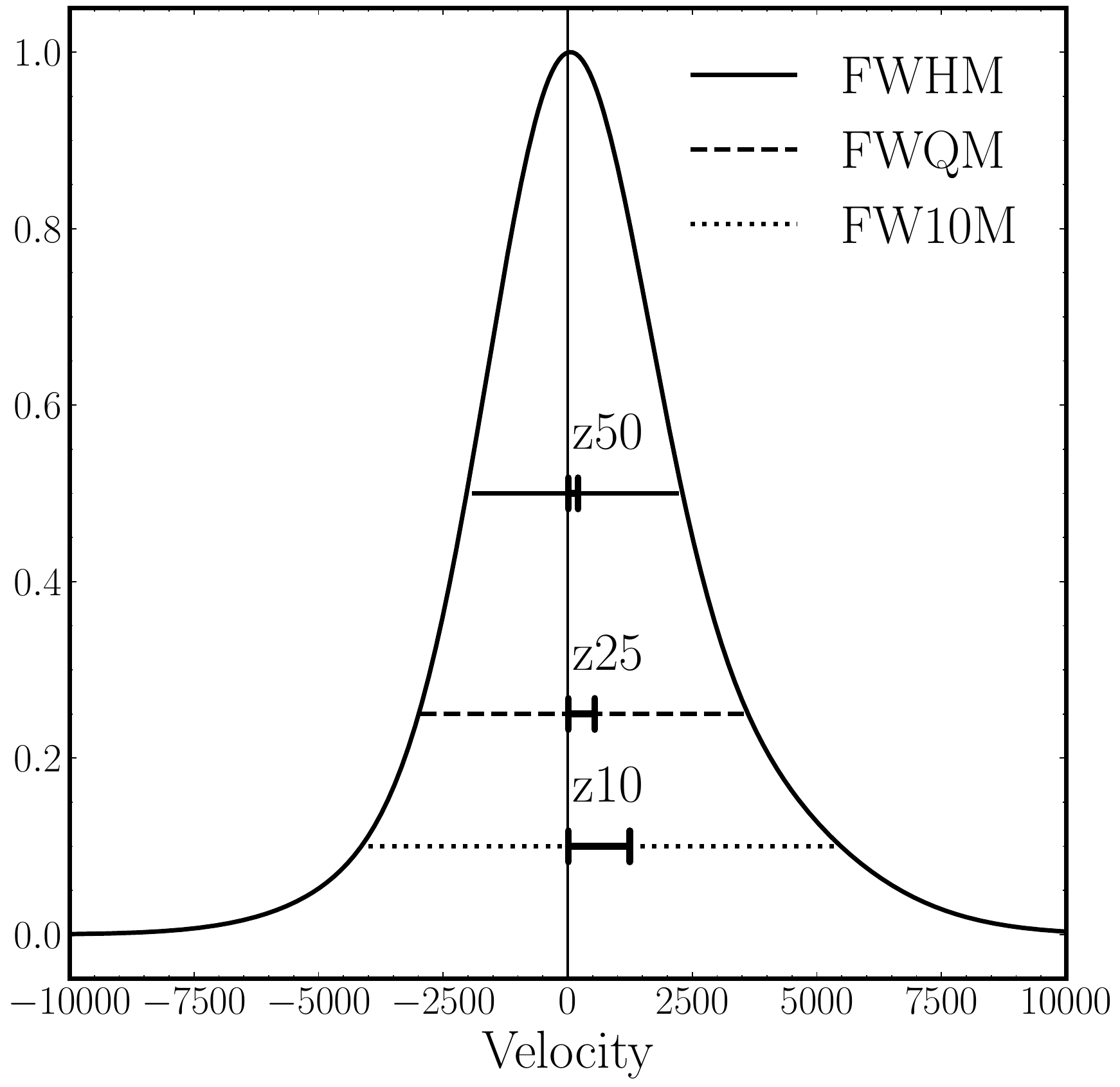}
\caption{Normalized broad line profile with indicated measured parameters: full widths at different level of maximum intensity - at 50\% FWHM, at 25\% FWQM, and at 10\% FW10M. The corresponding red asymmetries were also indicated with z50, z25, z10.}
\label{fig:anotate}
\end{figure}

Once the broad H$\alpha$ and H$\beta$ line profiles have been extracted, we measured their following properties (Figure \ref{fig:anotate}). The full width at different level of maximum intensity are: i) full width half maximum  - FWHM, ii) full width quarter maximum - FWQM, and iii) full width at 10\% maximum - FW10M. The asymmetry of the line, which are indicator of the intrinsic gravitational redshift (see Section 3 for details), is measured as a difference between the centroid of the line at different levels of intensity and position of the peak of the broad line. Thus we have also measured corresponding three asymmetries at 50\%, 25\%, and \%10 of the maximum intensity, denoted as z50, z25, z10, respectively (Figure \ref{fig:anotate}).
 We also measured the emission line dispersion $\sigma$ calculated as the second moment of the broad line profile using \cite{peterson2004} formalism (see their Equations 4 and 5).  Line widths and asymmetries were measured from the modeled broad line profiles, whereas the line dispersion was measured from the clean broad line profile, which is obtained from the observed spectra from which the contribution of the underlying continuum and satellite lines has been subtracted. The uncertainties are estimated as 1$\sigma$ errors derived from random subsets of N spectra \citep[as in e.g.,][]{kollat2011}.

In addition, we measured the broad line fluxes, as well as the Fe II optical emission in the range 4435–4685 \AA \, to get the $R_{\rm Fe II}$ parameter. The AGN continuum flux at 5100\AA \, was measured as a median of the integral in 5080-5120 \AA \, range from the reconstructed pure AGN spectra (Figure \ref{fig:host}). The luminosity was calculated using the luminosity distance obtained from the redshift and adopted cosmology (see Section 1). 

Table \ref{tab:par} lists the all measured spectral parameters, that is: the object ID, redshift, continuum luminosity L$_{5100\angstrom}$, optical Fe II emission R$_{\rm Fe II}$, full width and asymmetries at different level of maximum intensity for H$\alpha$ and H$\beta$ broad line. The full table is available online as supporting information.

\begin{table*}
\caption{Measured spectral parameters for the sample of 946 SDSS type 1 AGN. Columns are: SDSS object ID, redshift, continuum luminosity $\log \lambda L_{5100\angstrom}$, optical Fe II emission R$_{\rm Fe II}$, full width at half (FWHM), quarter (FWQM), and 10\% (FW10M) of maximum intensity, and corresponding asymmetries (z50, z25, z10). Line widths and asymmetries are listed for both H$\alpha$ and H$\beta$ broad line. The full table is available online as supporting information.}
\label{tab:par}
\resizebox{\textwidth}{!}{%
\begin{tabular}{lccccccccccccccc}
\hline
\multirow{3}{*}{SDSS ID} & \multirow{3}{*}{z} & $\lambda L_{5100\angstrom}$ & R$_{\rm Fe II}$ & \multicolumn{2}{c}{FWHM} & \multicolumn{2}{c}{FWQM} & \multicolumn{2}{c}{FW10M} &\multicolumn{2}{c}{z50}&\multicolumn{2}{c}{z25}&\multicolumn{2}{c}{z10}\\
 &  & [erg s$^{-1}$] &  & \multicolumn{2}{c}{[$\rm kms^{-1}$]} & \multicolumn{2}{c}{[$\rm kms^{-1}$]} & \multicolumn{2}{c}{[$\rm kms^{-1}$]} &\multicolumn{2}{c}{[$\rm km s^{-1}$]}&\multicolumn{2}{c}{[$\rm km s^{-1}$]}&\multicolumn{2}{c}{[$\rm km s^{-1}$]}\\
&  &  & &H$\beta$&H$\alpha$ &H$\beta$&H$\alpha$&H$\beta$&H$\alpha$&H$\beta$&H$\alpha$&H$\beta$&H$\alpha$&H$\beta$&H$\alpha$\\\hline
SDSS J145824.46+363119.5 & 0.25 & 44.51 & 0.68 & 2120 & 2100 & 3800 & 3200 & 6860 & 5940 & 0 & 0 & 150 & 0 & 150 & 0 \\ 
SDSS J095302.64+380145.2 & 0.27 & 44.55 & 0.25 & 3210 & 3020 & 5550 & 4620 & 8320 & 7220 & 290 & 0 & 1020 & 140 & 1460 & 730 \\ 
SDSS J004222.18-055823.4 & 0.2 & 44.02 & 0.43 & 3650 & 2700 & 5250 & 4300 & 6860 & 6210 & 0 & 0 & 0 & -180 & 0 & -270 \\ 
SDSS J142245.78+630739.1 & 0.16 & 43.94 & 0.59 & 3070 & 2240 & 5110 & 3570 & 7450 & 5530 & 290 & 0 & 730 & 90 & 1020 & 230 \\ 
SDSS J103208.42+405508.8 & 0.4 & 44.89 & 0.56 & 4160 & 3890 & 7080 & 6030 & 10370 & 9140 & 360 & 0 & 1240 & 180 & 1750 & 370 \\ 
SDSS J225603.37+273209.5 & 0.36 & 45.25 & 0.22 & 5460 & 4540 & 7800 & 7530 & 9980 & 10560 & 0 & 2520 & 0 & 3210 & 0 & 3210 \\ 
SDSS J032559.97+000800.7 & 0.36 & 44.8 & 0.11 & 6920 & 6850 & 9830 & 9820 & 12670 & 12650 & 0 & 460 & 0 & 500 & 0 & 500 \\ 
SDSS J140839.00+630600.5 & 0.26 & 44.51 & 0.18 & 5250 & 5120 & 7810 & 7500 & 10650 & 10290 & 0 & 180 & 0 & 460 & 0 & 870 \\ 
SDSS J224113.54-012108.8 & 0.06 & 43.21 & 0.09 & 6740 & 5760 & 9740 & 8140 & 12600 & 10560 & 2640 & -90 & 2710 & -90 & 2780 & -140 \\ 
SDSS J105007.75+113228.6 & 0.13 & 44.51 & 0.34 & 2190 & 2330 & 3650 & 3560 & 6780 & 5850 & 0 & 0 & 290 & 0 & 950 & 370 \\ 
  
\hline
\end{tabular}}
\end{table*}

\begin{figure}
    \centering
    \includegraphics[width=\columnwidth]{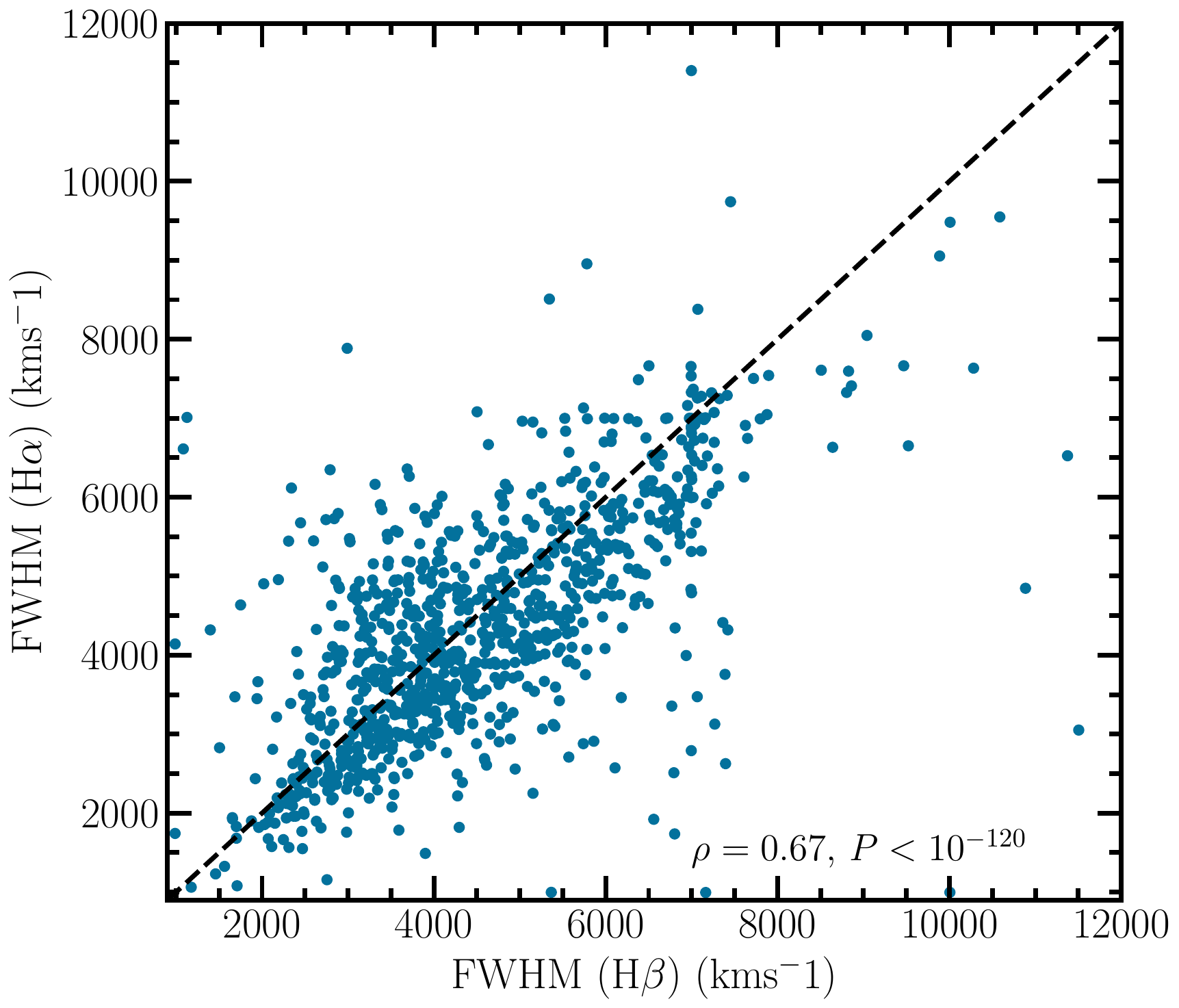}
    \caption{FWHM of H$\beta$ vs. FWHM of H$\alpha$ broad line for the total sample of type 1 AGNs. The one-to-one ratio is shown as dashed line. Pearson correlation coefficient together with corresponding P-value is also given.}
    \label{fig:fwhm}
\end{figure}

\begin{figure}
    \centering
    \includegraphics[width=0.95\columnwidth]{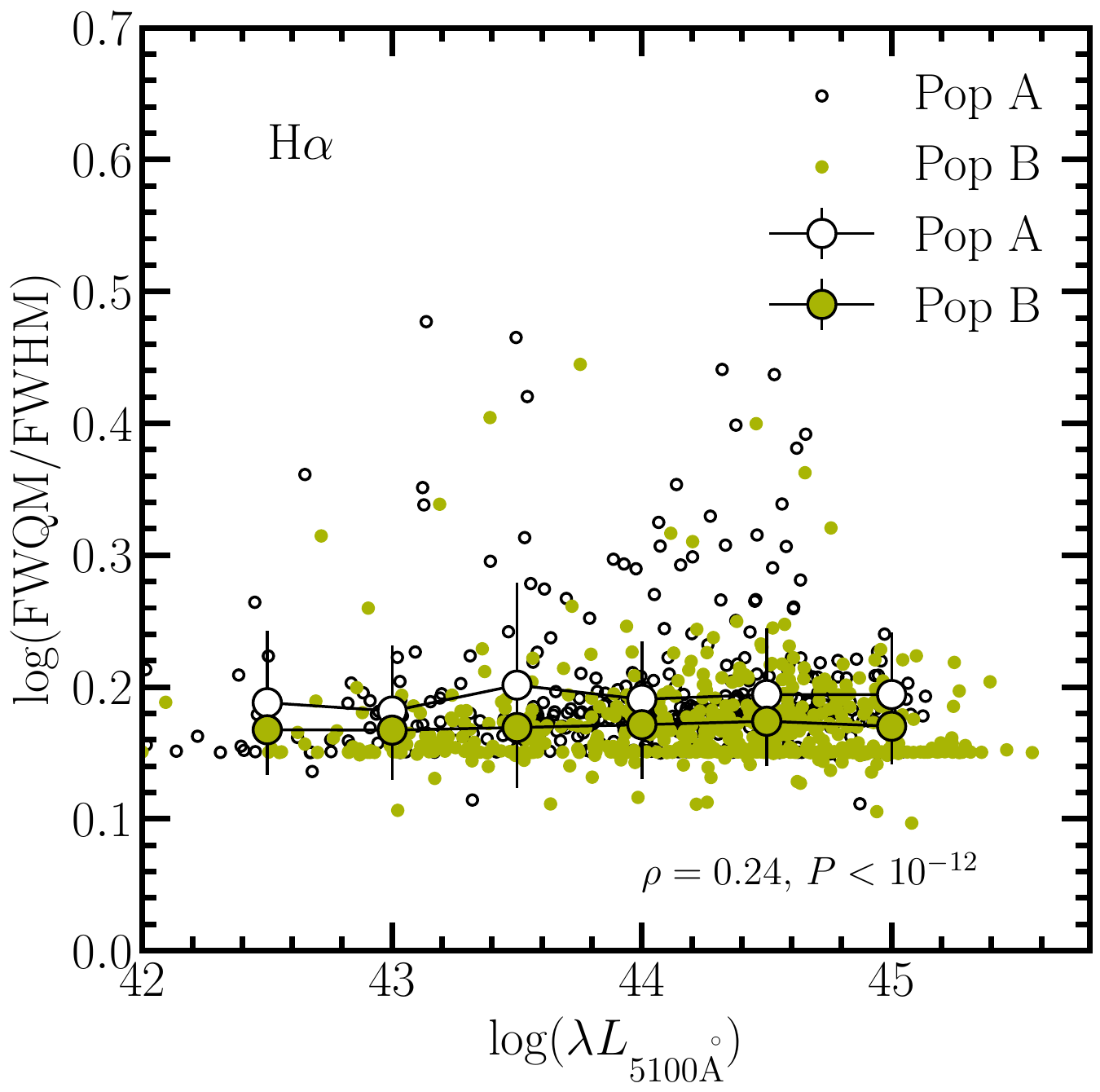}
    \includegraphics[width=0.95\columnwidth]{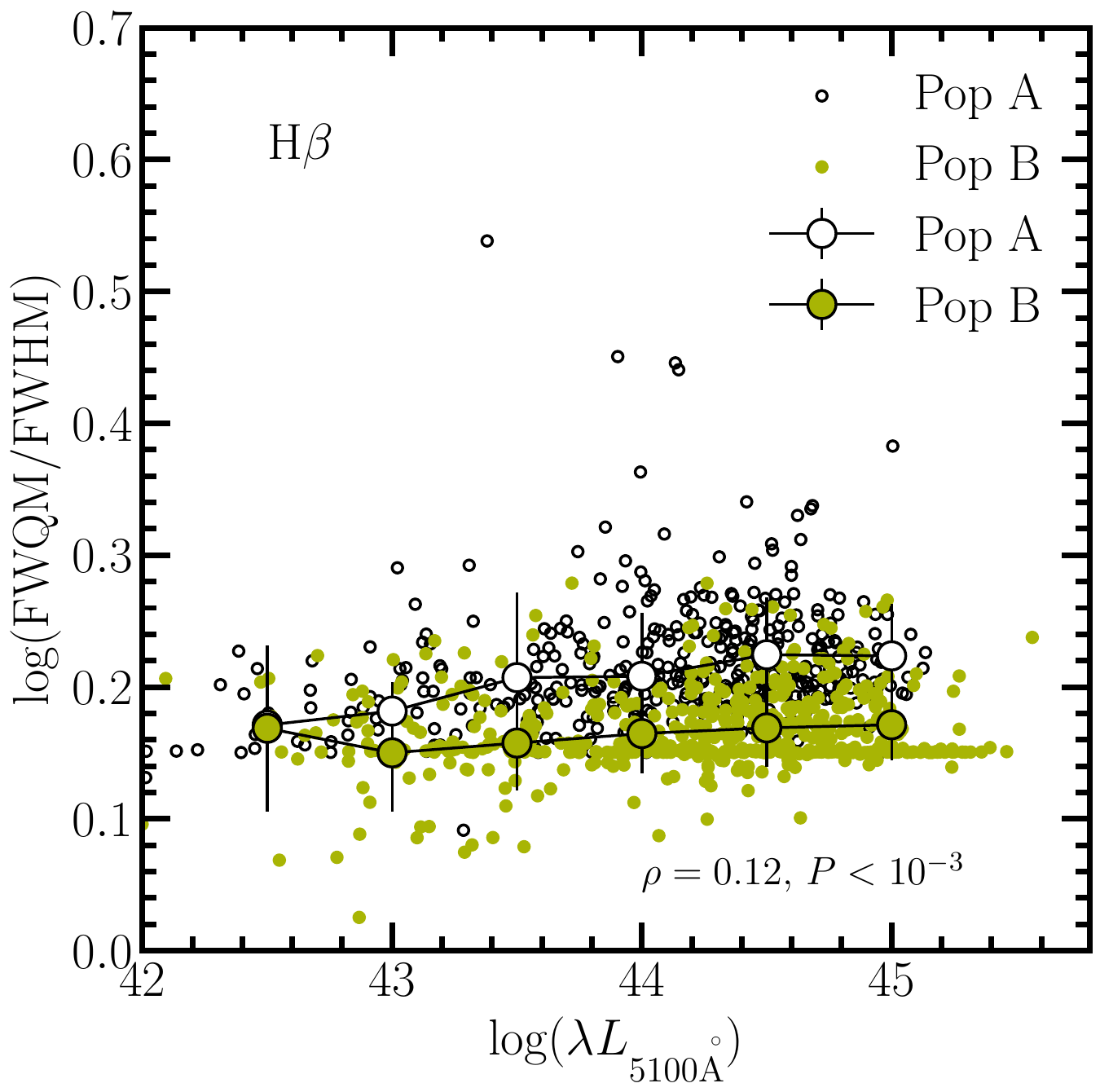}
    \caption{Ratio of the FWHM to FWQM vs. $\log(L_{5100\angstrom})$for H$\alpha$ (upper panel) and H$\beta$ (bottom panel) for the population A (open circles) and population B (green circles) sub-samples. The average values for aggregated continuum luminosity bins (0.5 dex) are also given (larger circles with error-bar).  Pearson correlation coefficient $\rho$ together with corresponding P-value is denoted on each plot.}
    \label{fig:ratio}
\end{figure}

\section{Kinematics and the virialization assumption of the BLR}
\label{sec:theory}

To investigate if the kinematics of the broad-line emitting gas is primarily driven with the SMBH gravitational force, that is to study the virialization of the BLR, we start from the following arguments:
\begin{enumerate}
    \item the ratio of the full line widths at different levels of line intensities should be constant across the luminosity scale;
    \item the gravitational redshift, here considered as an intrinsic shift, should be related to the square of the full line widths at different levels of line intensities.
\end{enumerate}

These arguments are derived from the following.  \cite{2019MNRAS.484.3180P} discussed that different region of the gas clouds from the BLR are contributing to the different part of the emission line \citep[see also][for a review]{2000ARA&A..38..521S}. Thus, emission of the regions closer to the SMBH will contribute to the line wings, whereas further regions will dominate the core of the emission line. Emission line widths can be used as a measure of the gas velocity. One can assume that widths at various scale of the line intensity, actually represent velocity of the different regions in the gas cloud, i.e. broader widths at lower levels of line intestines are indicators of the gas clouds closer to the SMBH. 

According to the virial theory we have \citep[see e.g. recent review by][and reference therein]{pop20}:
\begin{equation}
\label{eq:2}
    M_{\rm BH} = f_{\text{line}} \frac{\text{FWHM}^2R_{\rm line}}{G},
\end{equation}
where $M_{\rm BH}$ is mass of the black hole, $f_{\text{line}}$ is the virial factor accounting for the inclination and geometry of the BLR, $R_{\text{line}}$ is the distance to the line-emitting region, and $G$ is gravitational constant.
Using Eq. \ref{eq:2}, we can write that:
\begin{align}
M_{\rm BH} = f_{\text{line}} \frac{\text{FWHM}^2R_{\rm line, 1/2}}{G},\label{eq:3.1}\\
M_{\rm BH} = f_{\text{line}} \frac{\text{FWXM}^2R_{\rm line, X}}{G}\label{eq:4.1},
\end{align}
where $R_{\rm line, 1/2}$ is the distance to the line-emitting gas with observed velocities measured with $\rm FWHM$, whereas $R_{\rm line, X}$ is distance from the SMBH to the gaseous clouds with observed velocity indicated by the full width at some arbitrary X percentage of the line maximum intensity ($\rm FWXM$). 
If we assume that the $f$-factor for the emission-line
profiles measured at the half and X percentage of the maximum intensity are the same, combining Eqs. \ref{eq:3.1} and \ref{eq:4.1} we can write that:
\begin{equation}
\label{eq:vir}
    \frac{\text{FWXM}^2}{\text{FWHM}^2}=\frac{R_{\rm line, 1/2}}{R_{\rm line, X}}.
\end{equation} 
In addition, several reverberation mapping studies discovered an empirical relationship between the distance to the line-emitting region $R_\text{BLR}$ and central continuum luminosity $L_\text{cnt}$ \citep[see:][]{kaspi2000,bentz06}, which can be written in a form:
\begin{equation}
\label{eq:rl}
    \log R_\text{BLR}=a\cdot\log L_\text{cnt}+b.
\end{equation}
This relation should hold for the distance to different line-emitting regions. Thus, taking into the account Eqs. \ref{eq:vir} and \ref{eq:rl} one can expect that the logarithm of ratio of full widths at various intensity levels of the emission line should should be independent from the continuum luminosity. We will explore this assumption within our analysis to test if the virialization holds for the H$\alpha$ and H$\beta$ line-emitting regions.

\cite{zh90} argued that asymmetry of the H$\beta$ line wings can be explained with the gravitational redshift   \citep[see also e.g.,][etc.]{1977MNRAS.181P..89N,1995ApJ...447..496C, 2003A&A...412L..61K}.  The red asymmetry hes been discussed in terms of gravitation redshift in other lines as well, such as the UV Fe III lines \citep[][]{2018ApJ...862..104M}.  \cite{2020ApJ...903...44P} also report the strong red asymmetry in UV Mg II and C IV lines of a sample of blazars, which was discussed in terms of gravitational origin.

For the emitting gas at a distance $R$ from the SMBH, where $M_{\rm BH}$ is the mass of the black hole,  in the weak field approximation
the combined gravitational and transverse Doppler redshift will tend to \citep[see e.g.,][]{2015Ap&SS.360...41B,liu17, 2018ApJ...862..104M, 2019MNRAS.484.3180P,2022ApJ...928...60L}:
\begin{equation}
\label{eq:3}
    z_G = \frac{3}{2} \frac{GM_{\rm BH}}{c^2R}
\end{equation}
where $c$ is speed of the light.  Note that the transverse Doppler redshift being inversely proportional to the Lorentz factor, is entirely a special relativistic effect but, being independent from orientation, cannot be distinguished from the gravitational contribution. Previous analysis of AGN spectra suggested that the broad line redshifts could be mainly the result of gravitational redshift \citep[see e.g. discussion in][and references therein]{2018ApJ...862..104M}.

If one assume that the radius $R$, in Eqs. (\ref{eq:2}) and (\ref{eq:3}), represent the same photometric radius, it is expected to have:
\begin{equation}
\label{eq:4}
    z_G\sim\text{FWHM}^2.
\end{equation}
The asymmetry of the line, thus the gravitational redshift, can be quantified as a difference between the centroid of the line at different levels of intensity and position of the peak of the broad line \citep[for more explanation see:][]{jon}. Accordingly, if the broad-line emitting gas is virialized, one can expect to have linear relationship between logarithms of $z_G$ and FWHM of the line \citep[see also][]{2019MNRAS.484.3180P}. We use also this to explore the virialization in the BLR.

 We test the kinematics of the BLR emitting H$\alpha$ and H$\beta$ emission lines by exploring relation between ratio of FWHM/$\sigma$ vs. FWHM. Especially, because the line dispersion $\sigma$ is generally more sensitive to the line wings and less to the line core {\citep{collin2006, kollat2011}}. The rato of FWHM/$\sigma$ depends on the line profile: for gaussian profile we have FWHM/$\sigma$ = 2.35, for rectangular function ratio is 3.46, while for Lorentzian profiles  FWHM/$\sigma \rightarrow 0$ \citep[see e.g][]{kollat2013}. According to \cite{collin2006} and also \cite{kollat2011} broad lines have tendency to have flat-topped profiles, whereas narrower lines have more extended wings. \cite{collin2006} divided their sample of AGN into two populations based on the line profile: $\rm FWHM/\sigma_{\rm line}<2.35$ (population 1) and $\rm FWHM/\sigma_{\rm line}>2.35$ (population 2), finding that these basically correspond to separation to population A and B suggested by \cite{2000ARA&A..38..521S}. Instead, \cite{kollat2011, kollat2013} showed that there is more a continuous transition from narrow to broad-line objects. We explore these findings in our sample.

\begin{figure}
    \centering
    \includegraphics[width=0.95\columnwidth]{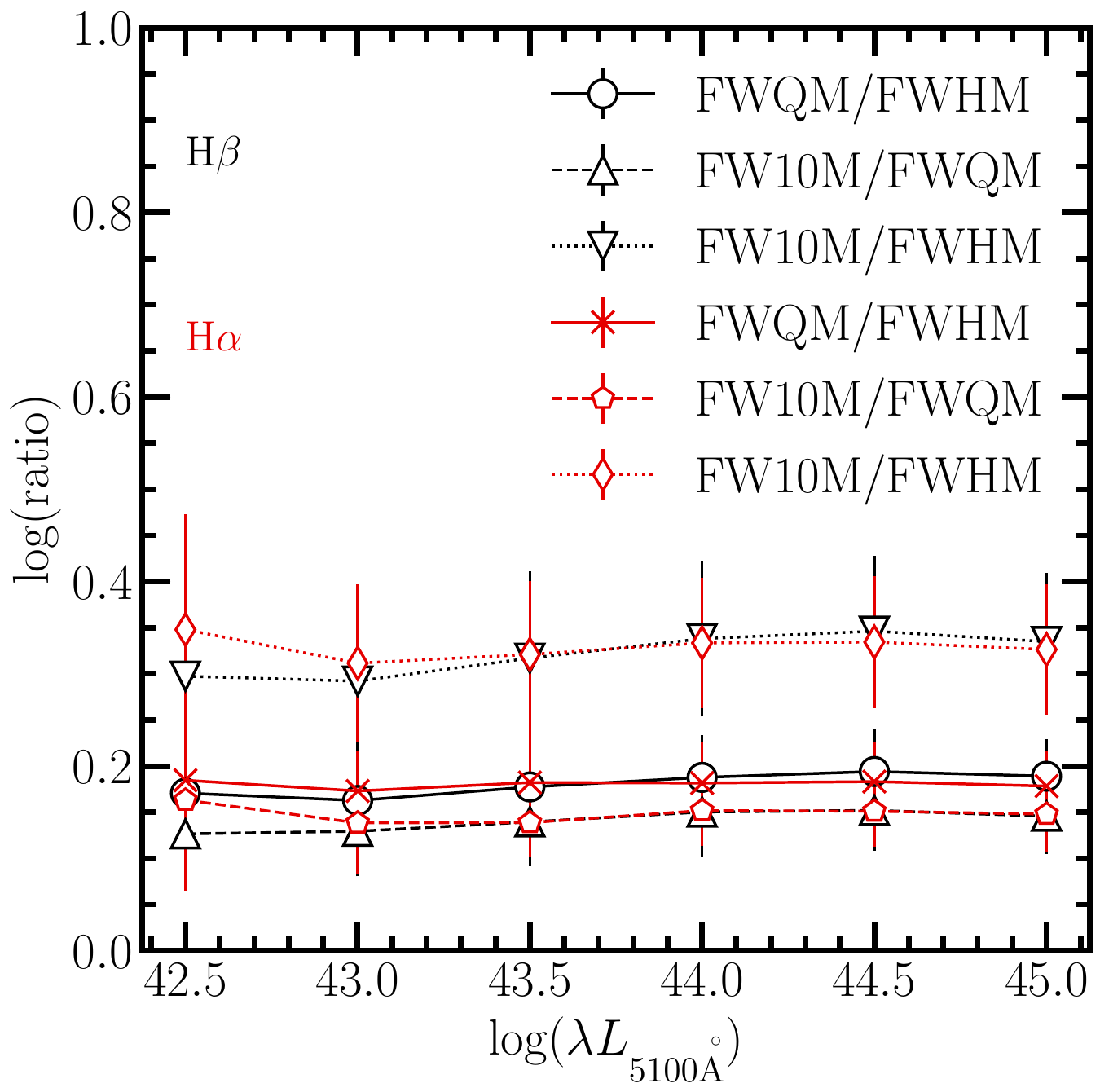}
    \caption{Logarithms of different line width ratios (denoted in upper right corner) across the luminosity scale. The values are averaged within continuum luminosity bins of 0.5 dex.}
    \label{fig:ratio_all}
\end{figure}

   \begin{figure}
    \centering
    \includegraphics[width=\columnwidth]{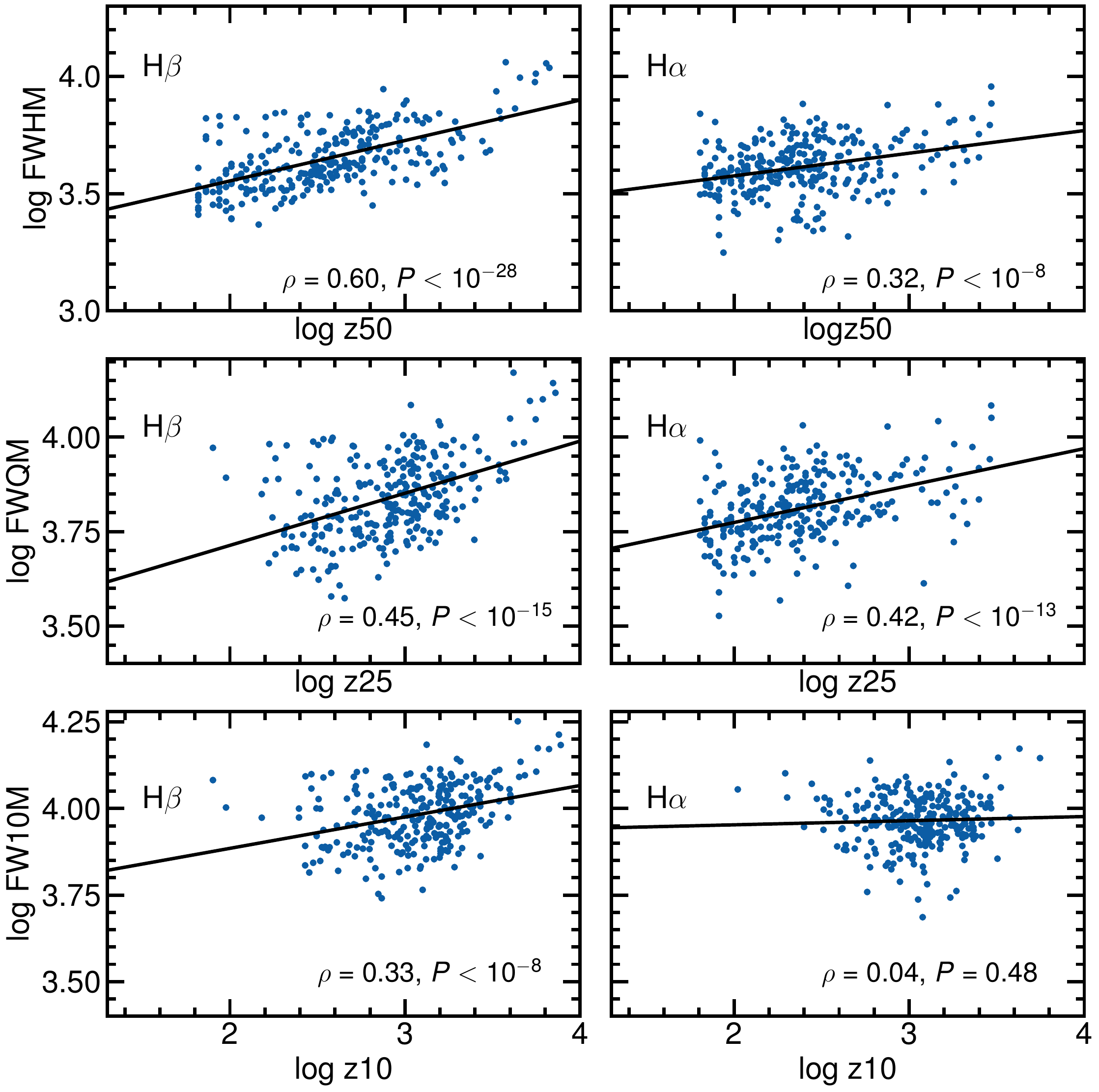}
    \caption{Correlation between asymmetry (i.e. gravitational redshift) measured at 50 (top panel)s, 25 (middle panels), and 10 (bottom panels) percent of line intensity versus corresponding full widths of the broad H$\beta$ (left), and H$\alpha$ line (right). Best linear fitting is given as a solid line. Pearson correlation coefficient, together with P-value is denoted on each plot.}
    \label{fig:zfwhm}
\end{figure}

\section{Results \& Discussion}

For the sample of 946 SDSS type 1 AGNs, we analyze in more details the measured H$\alpha$ and H$\beta$  broad line parameters, i.e., full widths and asymmetries at different levels of maximum intensities. 

In Figure \ref{fig:fwhm} the FWHM of H$\beta$ vs. FWHM of H$\alpha$ broad line is plotted, together with the one-to-one line to guide the eye. It is clear that the H$\alpha$ and H$\beta$ line widths are similar and well correlated (Pearson correlation coefficient $\rho=0.67$), supporting that similar kinematics of their emitting regions. 

\begin{figure}
    \centering
    \includegraphics[width=0.9\columnwidth]{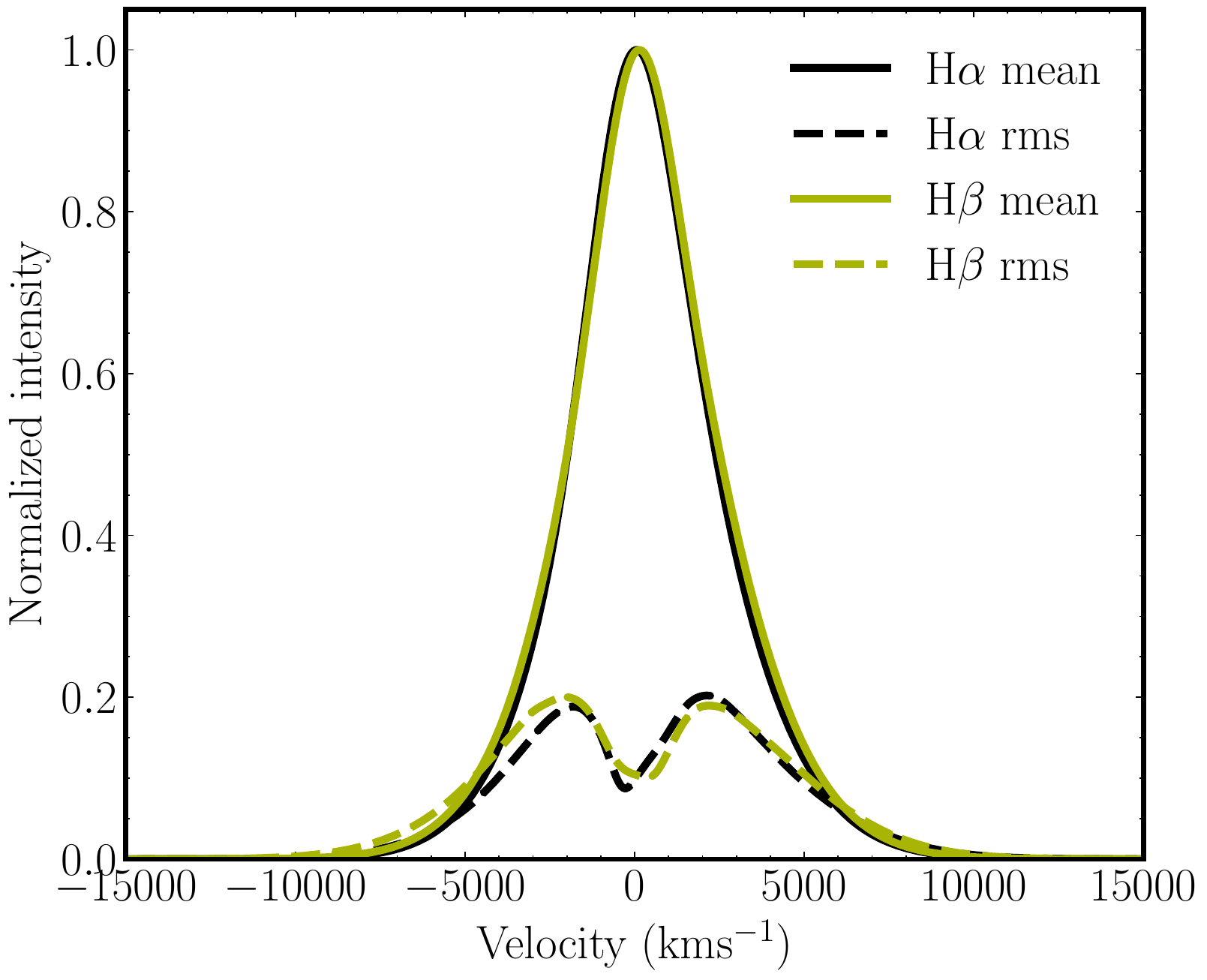}
    \includegraphics[width=0.9\columnwidth]{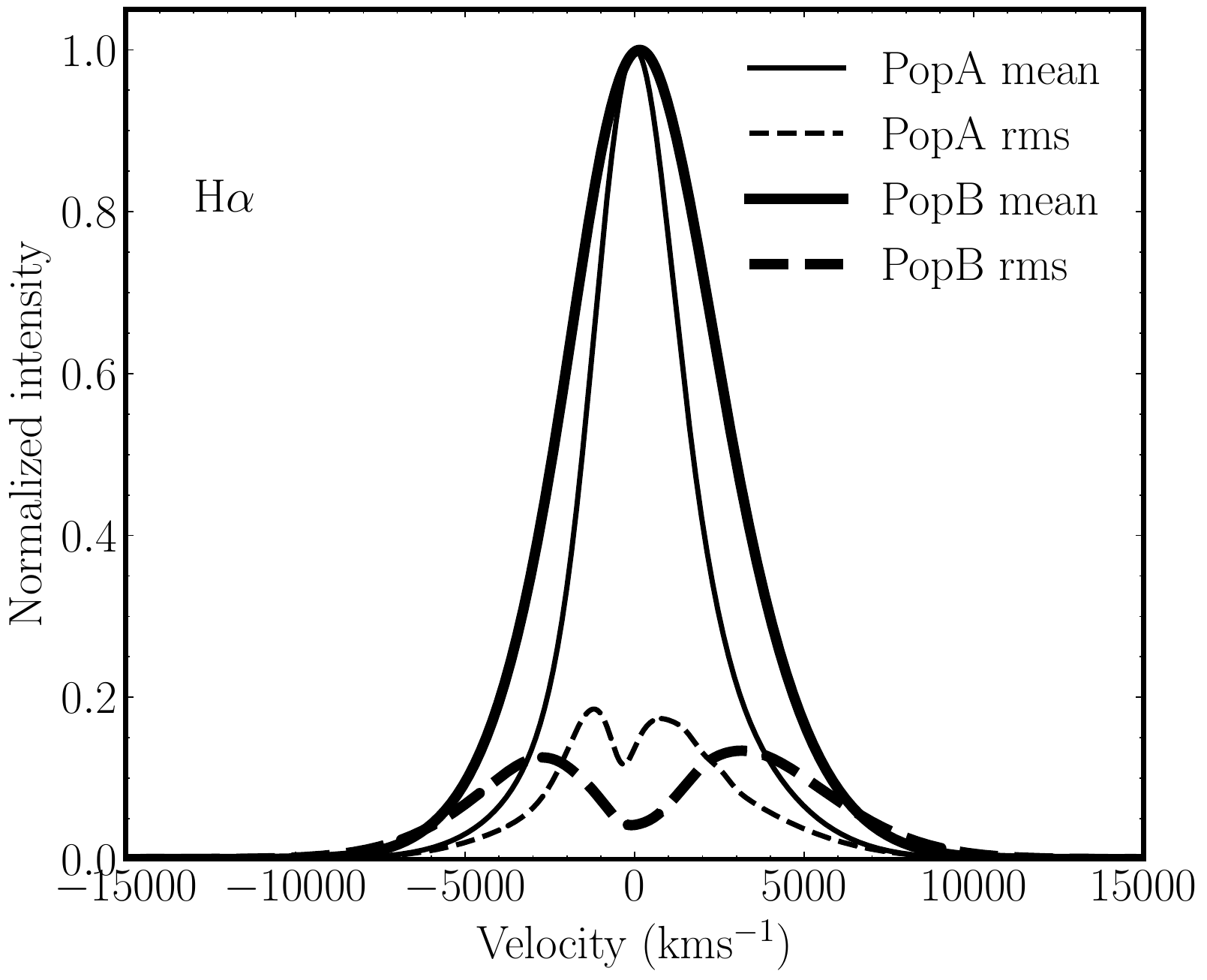}
     \includegraphics[width=0.9\columnwidth]{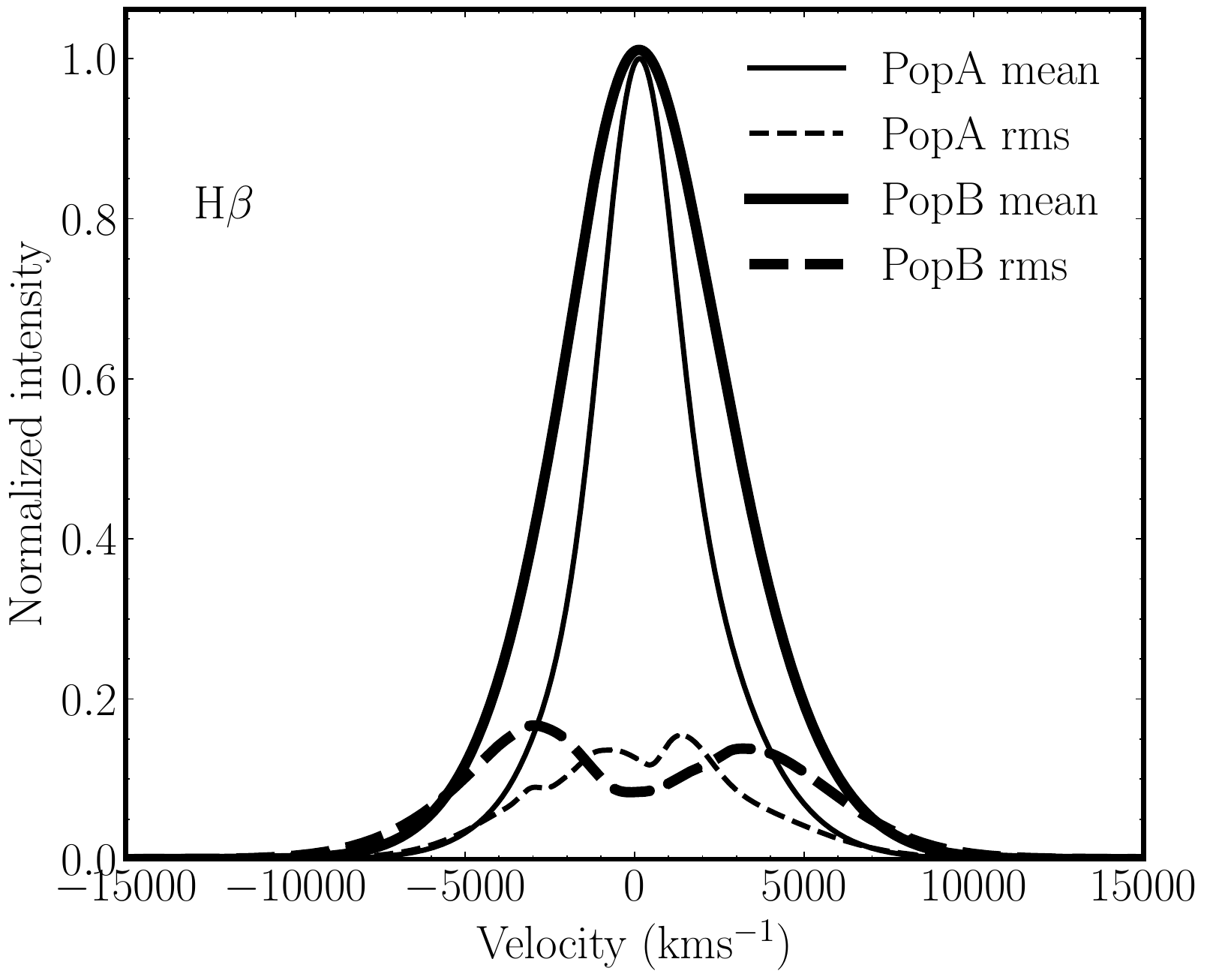}
    \caption{Upper: mean (full line) and rms (dashed line) of broad lines H$\alpha$ (black) and H$\beta$ (green)  of the total sample. Middle: mean (full line) and rms (dashed line) of broad H$\alpha$ for the Population A (thin black line)  and Population B (thick black line). Bottom: same as in the middle, but for the broad H$\beta$ line.}
    \label{fig:rms}
\end{figure}

In order to test whether arguments from Section \ref{sec:theory} are fulfilled, we explored relation between line widths and luminosity, as well as line widths and asymmetries. We start with the logarithm of the ratio of the full widths at quarter (FWQM) and half (FWHM) maximum intensity of both broad H$\beta$ and H$\alpha$ lines with the continuum luminosity at 5100$\angstrom$ (Fig. \ref{fig:ratio}).
Furthermore, we aggregate data into continuum luminosity bins (0.5 dex), starting with $\log(\lambda L_{5100\angstrom})$, and give their average values. We separately considered Population A and B sub-type of AGNs. From Fig. \ref{fig:ratio} it is straightforward to realise that widths ratio is rather constant across the luminosity scale for both H$\alpha$ and H$\beta$ line. This is supported with the very low level of correlation (Pearson correlation coefficients are 0.24 and 0.12 for H$\alpha$ and H$\beta$, respectively). The constant trend is even more emphasized if we take the data aggregated in luminosity bins. The deviation of the trend is noticeable in the lower continuum bins, that is probably due to small number of data used for aggregation. We tested this for the ratios of all other measured line widths (FWHM, FWQM, FW10) and presented in Figure \ref{fig:ratio_all}. The findings are the same, there is no trend seen in the ratios of line widths across different luminosities.

\cite{2019MNRAS.484.3180P} argued that contribution to different parts of the broad line are coming from different locations withing gaseous clouds surrounding the SMBH. Wings of the line are originating in clouds closer to the SMBH, whereas the line core is form the outer parts. If gas is virialized it is natural to assume that relation given in Eq. \ref{eq:4}  is held across all the regions contribution to the emission line. With all above, if the gas is virialized we expect that logarithm of ratios of full widths should be constant across different continuum luminosity. One can see from the Fig. \ref{fig:ratio} that our findings are in line with the virial theory, and the kinematics of H$\alpha$ and H$\beta$ broad-line emitting-region is driven by the SMBH. Furthermore, clearly there is no much difference if one separately observe Population A and B types of AGNs.

Further, we observe whether the second arguments, involving the gravitational intrinsic redshift (Section \ref{sec:theory}) is applicable for our sample. Following Eq.\ref{eq:4}, in Figure \ref{fig:zfwhm} we plot correlations between the red asymmetry (i.e., gravitational redshift) measured at 50 (top panels), 25 (middle panels), and 10 (bottom panels) percent of line intensity versus corresponding full widths of the broad H$\beta$ (left) and H$\alpha$ line (right). We exclude from the analysis, detected asymmetries below the SDSS instrumental resolution of 70 km s$^{-1}$.
Figure \ref{fig:zfwhm} reveals that the line widths at half and quarter maximum intensity are well-correlated with the corresponding asymmetries, being a measure of the intrinsic gravitational redshift. The correlations are particularly solid in case of FWHM of H$\beta$ and H$\alpha$, which is supported with the Pearson correlation coefficient of 0.60 and 0.45, respectively  (denoted on each plot n Figure \ref{fig:zfwhm}). The asymmetry measured in line wings is stronger, and show weaker correlation with the corresponding width (see FW10M vs. z10 in Figure \ref{fig:zfwhm}, bottom plots), being practically absent in case of H$\alpha$ line. These are indicating that possibly the region contributing to the line-wings, especially in H$\alpha$, is not virialized, i.e., there are possible presence of radial motions such as inflows/outflows \citep[][]{2019MNRAS.484.3180P}. We note that measurements in the line wings are most sensitive to the estimates of the underlying continuum, which is simultaneously fitted during this automated fitting procedure. 

\begin{figure}
    \centering
    \includegraphics[width=1.05\columnwidth, height=7cm]{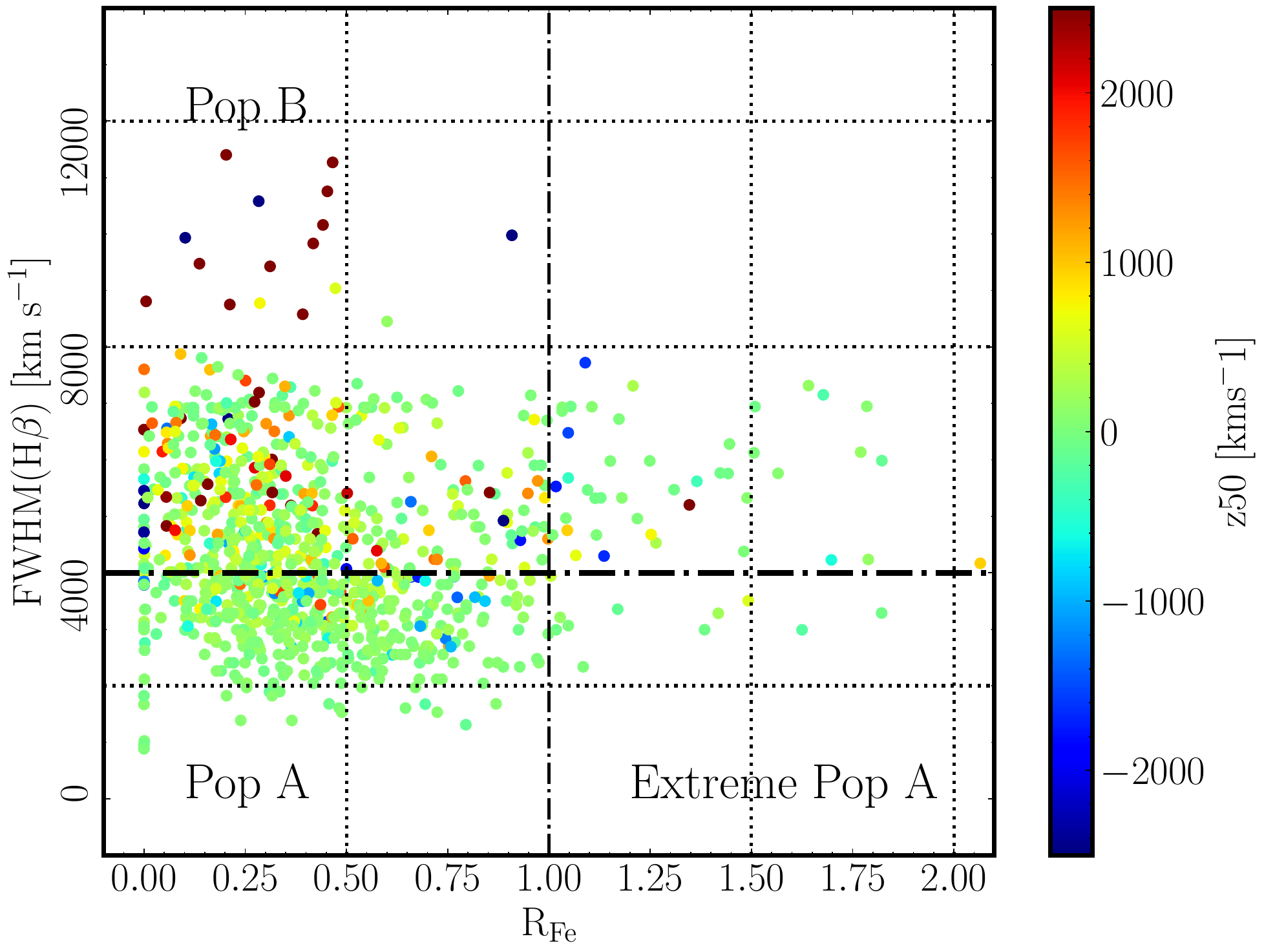}
    \caption{Location of the studied sample in the FWHM(H$\beta$) - R$_{\rm Fe II}$ plane (so-called AGN main sequence) showing red and blue asymmetry in the H$\beta$ line. Dot-dashed horizontal (FWHM(H$\beta$)=4000 km/s) and vertical (R$_{Fe II}$=1) line divide the Population B, Population A, and extreme population A. The colorbar indicates the asymmetry measured at half maximum z50 in km/s.}
    \label{fig:main}
\end{figure}

\begin{figure*}
    \centering
    \includegraphics[width=0.60\columnwidth]{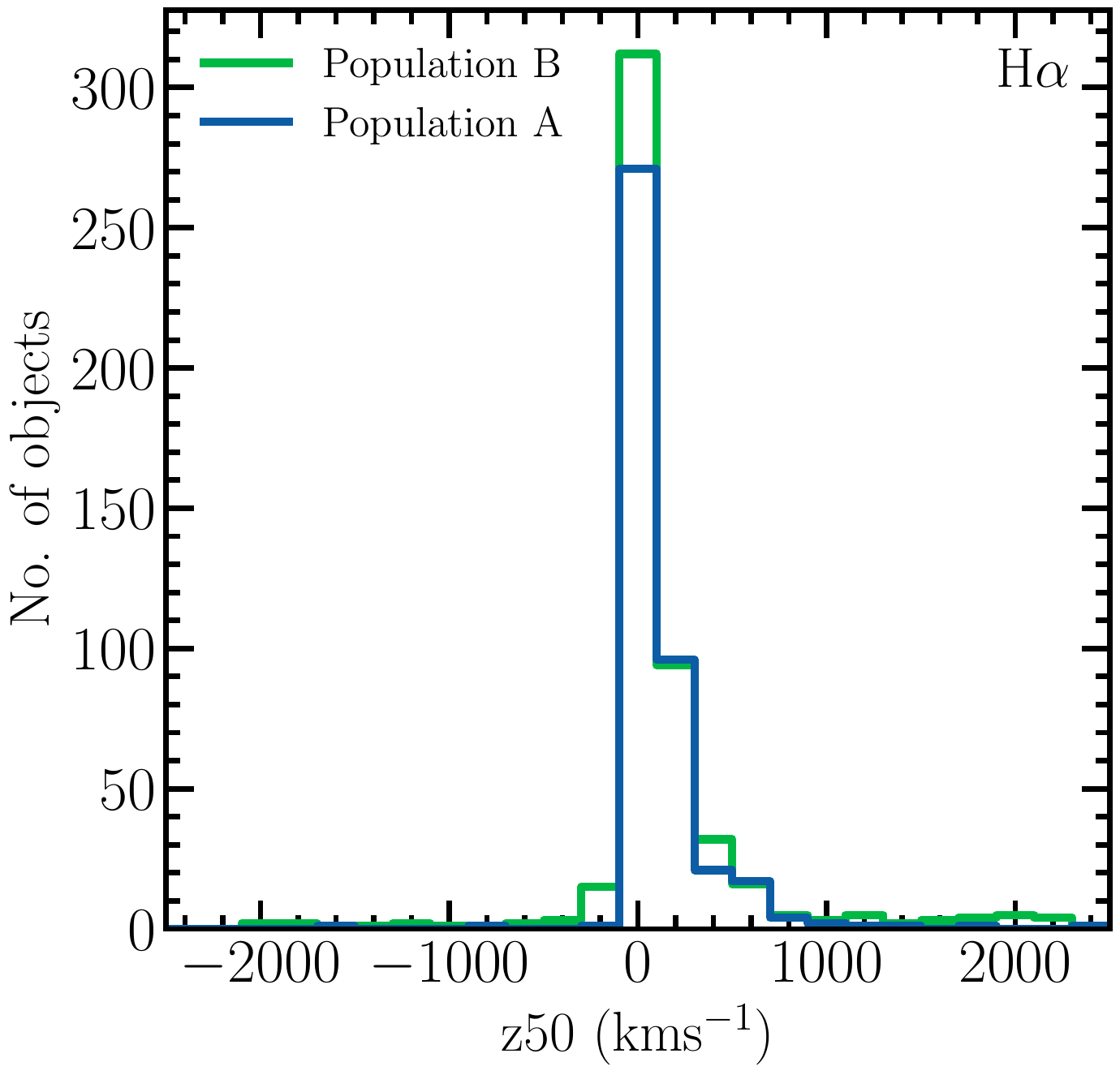}
    \includegraphics[width=0.60\columnwidth]{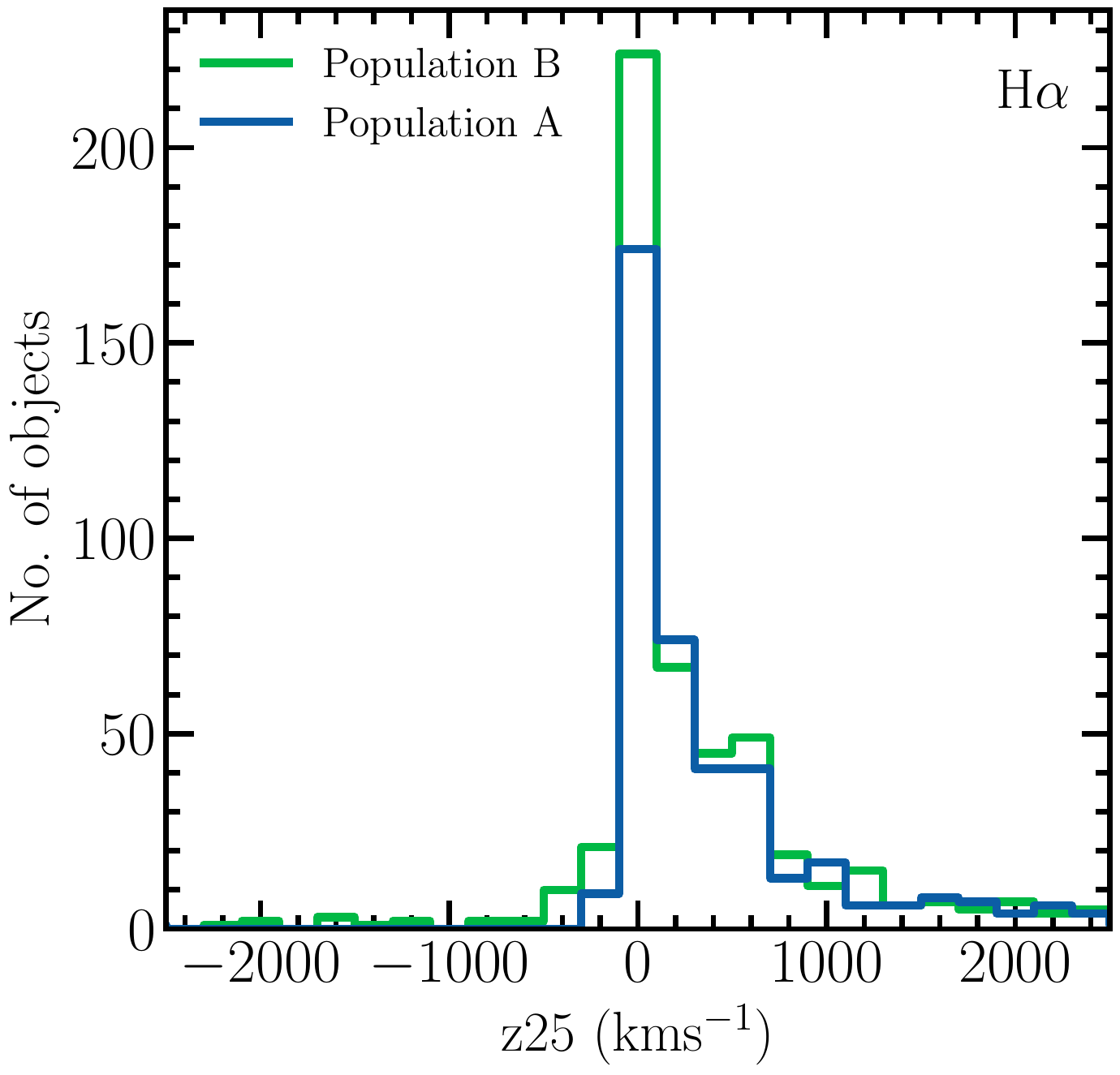}
    \includegraphics[width=0.60\columnwidth]{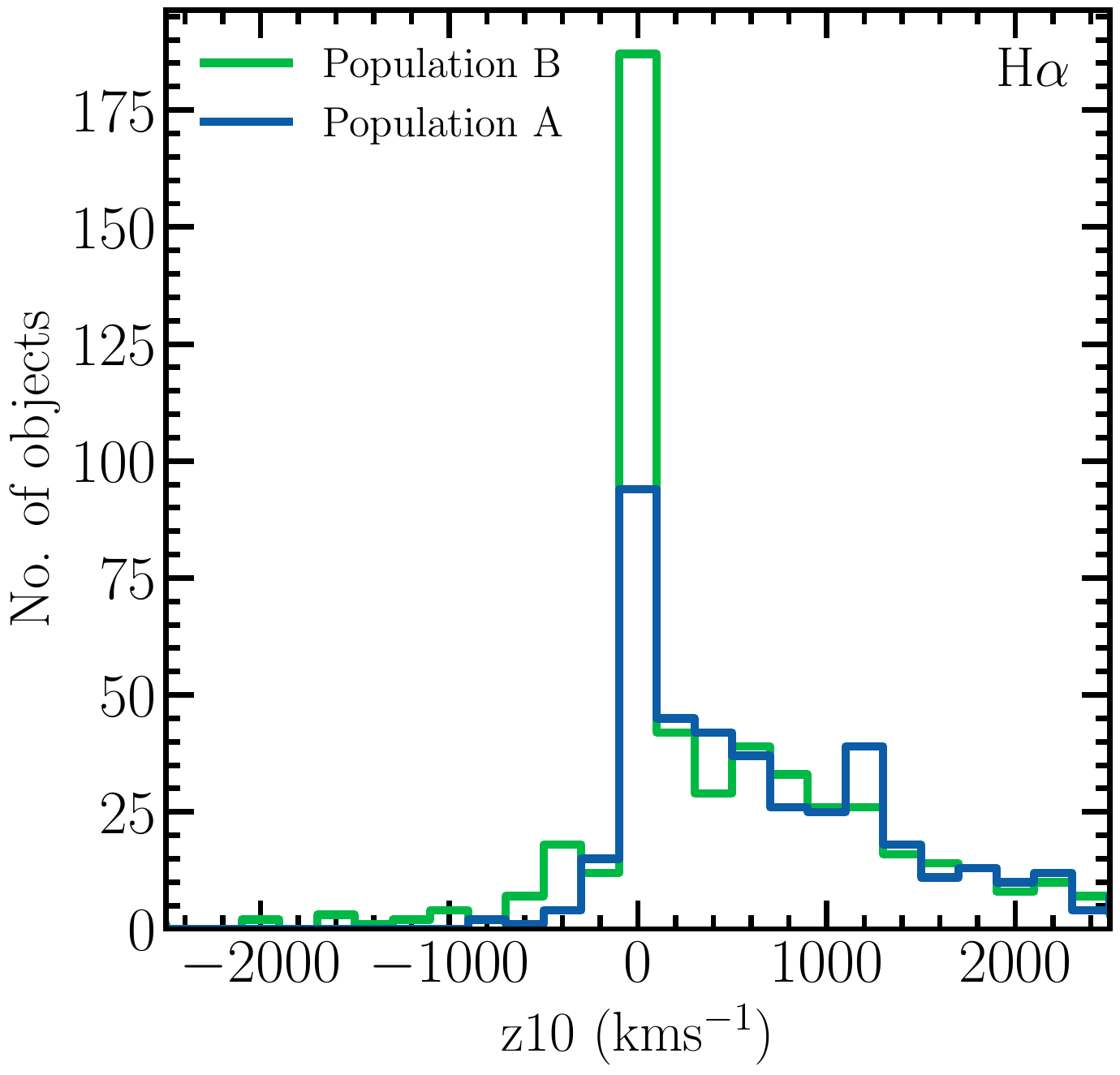}
    \includegraphics[width=0.60\columnwidth]{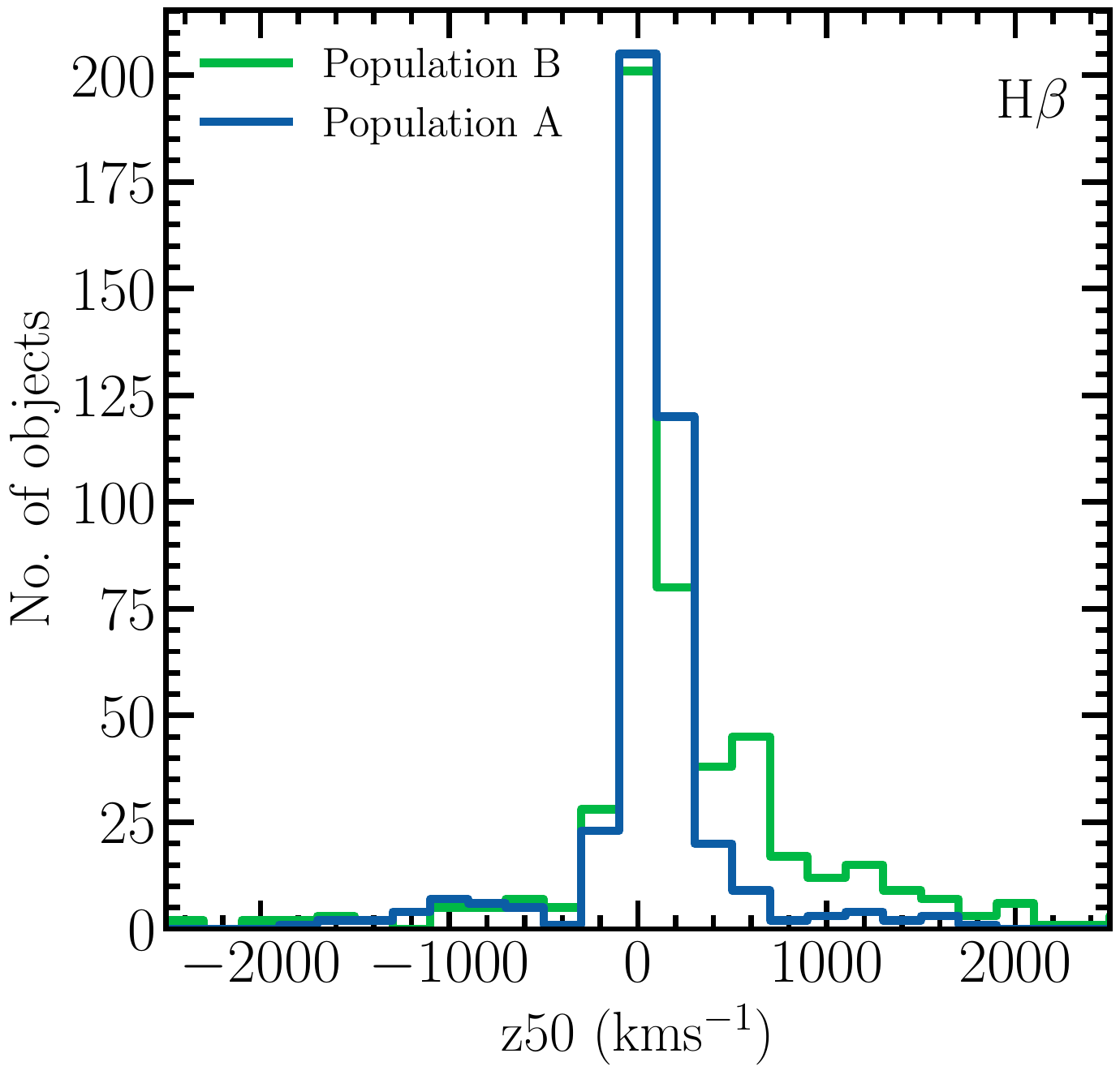}        \includegraphics[width=0.60\columnwidth]{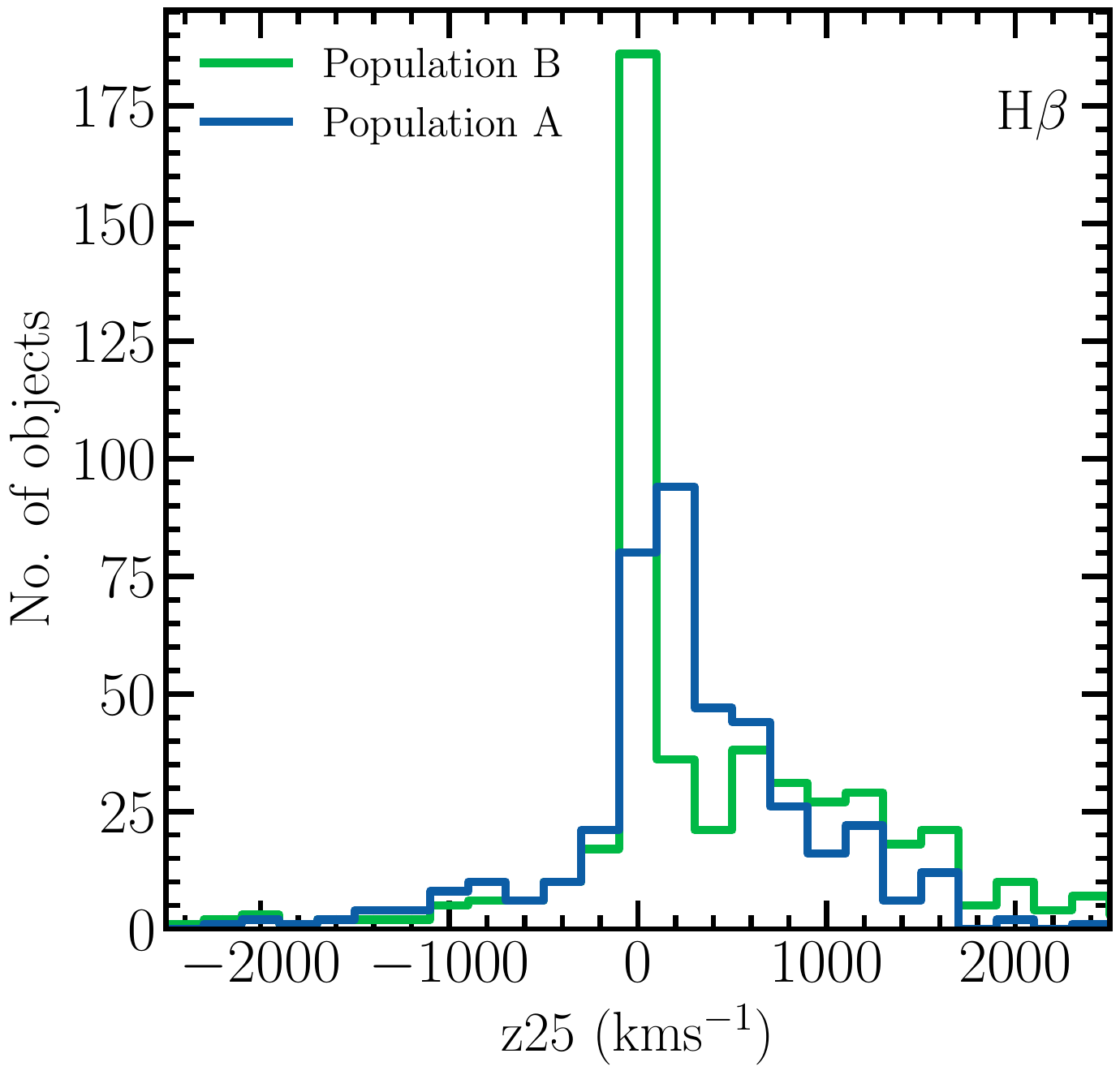}
    \includegraphics[width=0.60\columnwidth]{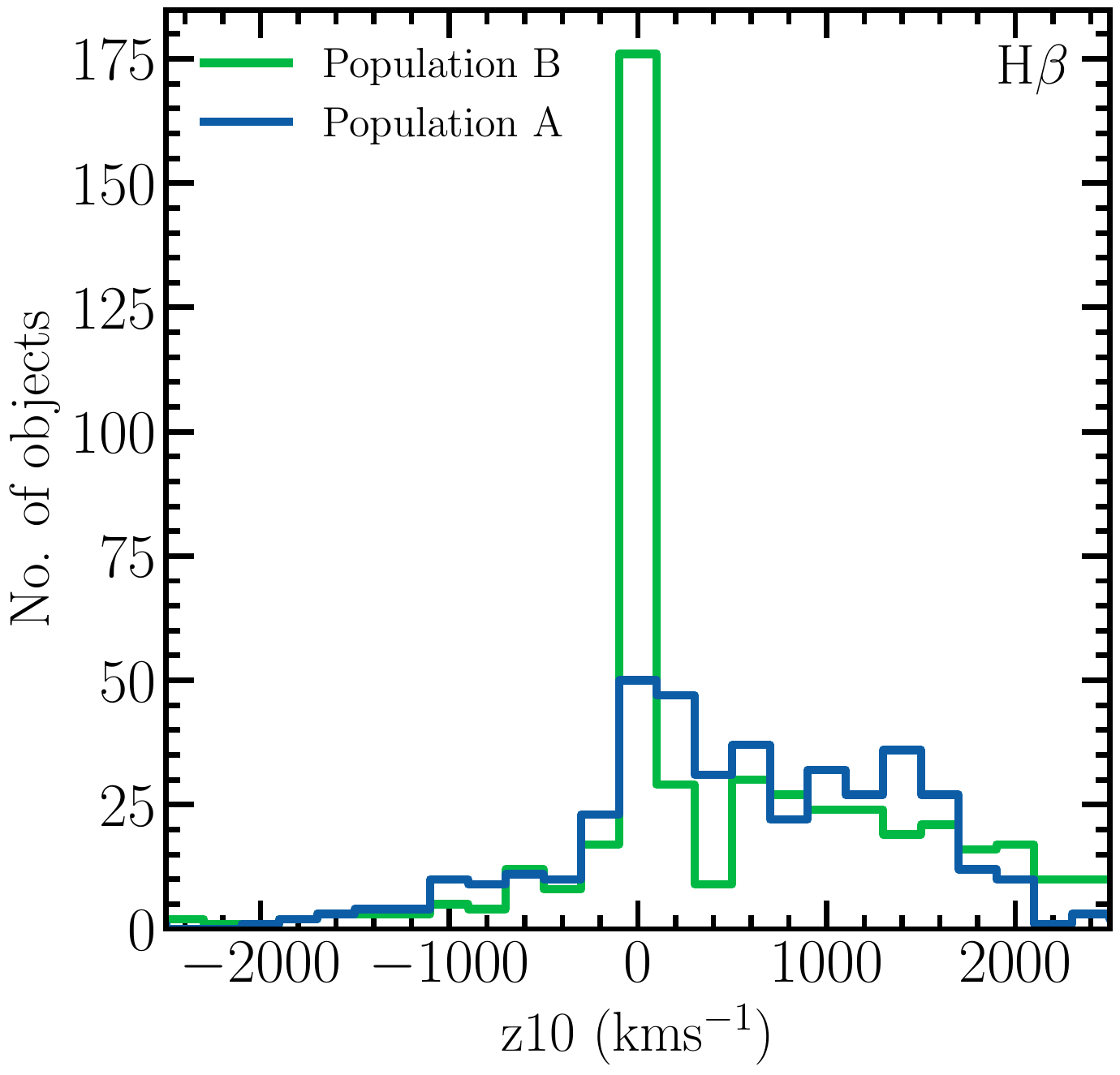}
    \caption{Histogram of the asymmetries measured at 50\% (left panels), 25\% (middle panels), and 10\% (right panels) 
of line intensity for the broad H$\alpha$ (upper panels) and
H$\beta$ line (bottom panels), for Population A (blue line) and Population B (green line) objects. }
    \label{fig:hist_z}
\end{figure*}

The mean and root-mean-square (rms) profiles of the normalized H$\alpha$ and H$\beta$ broad line are shown in Figure \ref{fig:rms}. Both mean profiles for the total sample (Figure \ref{fig:rms}, upper panel) are symmetric, and practically identical. The lack of asymmetry in H$\beta$ line was already noted by \cite{2019MNRAS.484.3180P} for H$\beta$ line. These authors also showed that in case of Mg II and H$\beta$ line, the largest difference is seen in the line wings. However, here we see no difference in the mean profiles of H$\alpha$ and H$\beta$ (Figure \ref{fig:rms}, upper panel).
We tested also is there any difference between the mean and rms profiles in population A and B (Figure \ref{fig:rms}, middle and bottom panels). As expected, population B mean profile is wider, but both are symmetric, with very subtle red asymmetry seen in H$\alpha$ line. 

All these results support that the BLR emitting Hydrogen Balmer lines is virialized, i.e., the gas motion is primarily driven by the SMBH gravity.

We tested how the asymmetry, measured in the H$\beta$ line at the 50\% of line maximum, differs across the AGN main sequence (Figure \ref{fig:main}). First we note that our sample covers nicely the known main-sequence elbow shape in the FWHM(H$\beta$) - R$_{\rm Fe II}$ plane \citep[][]{2000ARA&A..38..521S,2001ApJ...558..553M,2014Natur.513..210S}. In case of population A, on average, asymmetry does not depend on the position of the object in the FWHM(H$\beta$) - R$_{\rm Fe II}$ plane. However, population B object show presence of more asymmetric profiles (see colorbar in Figure \ref{fig:main}). \cite{2010MNRAS.403.1759Z} showed that the population B object with broader profiles show higher red asymmetry. 
 We present in Figure \ref{fig:hist_z} histogram of the asymmetries measured at 50\%, 25\%, and 10\% percent of line intensity for the broad H$\alpha$ and H$\beta$ line for both Population A  and Population B objects. The majority of object show no (or very weak) asymmetry as measured with z50 (left panels, Figure \ref{fig:hist_z}), with population B object showing more asymmetric profiles in H$\beta$ line (as also concluded from Figure \ref{fig:main}). When measuring the asymmetry in line wings (see middle and left panels, Figure \ref{fig:hist_z}), the number of object showing red asymmetry increases, especially in H$\beta$ line. However, this could be due to fitting poorly accounting for the Fe II contribution, as well as estimates of the underlying continuum, for which the line wings are the most sensitive. On the other hand, this is could be also due to the presence of radial motion, as already discussed. 

\begin{figure}
    \centering
    \includegraphics[width=\columnwidth]{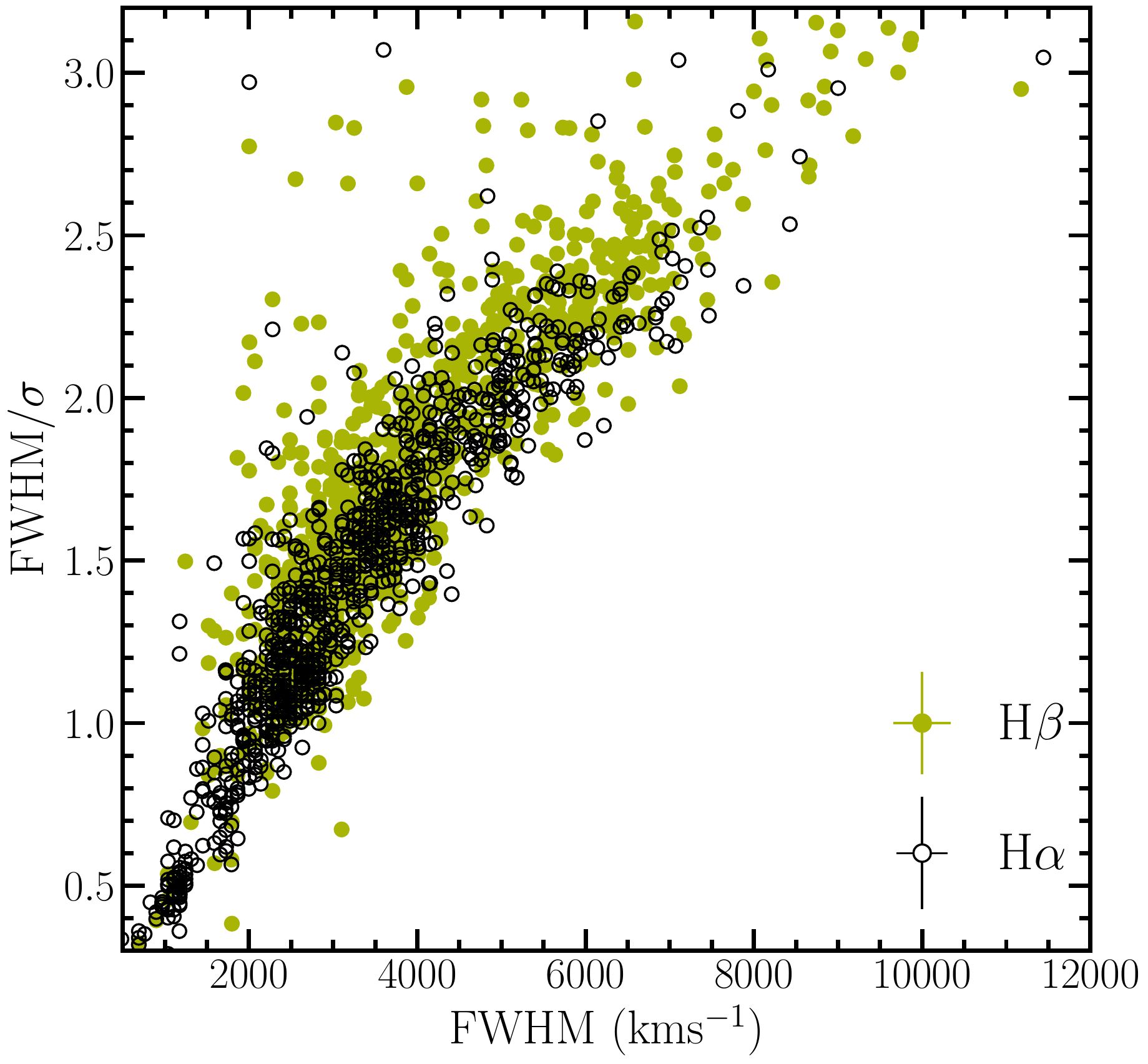}
    \caption{FWHM/$\sigma$ vs. FWHM for H$\alpha$ line (open circles) and for H$\beta$ line (green circles).  Mean error-bars are indicated on the plot.}
    \label{fig:sigma}
\end{figure}

 In the Figure \ref{fig:sigma} we presented the ratio of FWHM/$\sigma$ vs. FWHM for the H$\alpha$ (open circles) and H$\beta$ (green circles) broad lines. The results show that the broad line profiles vary systematically with the increase of the line widths in the same manner for both H$\alpha$ and H$\beta$. Profiles of broader lines tend to be more flat-topped, whereas the narrower profiles have more prominent line wings \citep[][]{collin2006, kollat2011}. 
 These findings are supporting the general trend found by \cite{kollat2011}, that the line-broadening is dominantly by rotation. We also see the same smooth transition between population A and B, both in H$\alpha$ and H$\beta$, supporting our previous discussion that they follow the same kinematics.

\section{Conclusions}   
 
In this work we study the kinematics of the BLR in a sample of 946 type 1 AGN selected from the SDSS DR16 only to have high S/N ratio and contain both H$\alpha$ and H$\beta$ emission lines. We perform careful extraction of the broad H$\beta$ and H$\alpha$ line from the pure AGN spectra, and measure their width and asymmetry. Our conclusions can be summarized as:
\begin{enumerate}
    \item the FWHM of H$\alpha$ and H$\beta$ are well correlated, indicating that the kinematics of the regions emitting these lines is the same. This is supported with symmetric mean H$\alpha$ and H$\beta$ broad line profiles, being almost identical; 
    \item the ratio of the full widths at quarter and half maximum intensity of H$\alpha$ and H$\beta$ lines are remaining the same across the luminosity scale. This applies also for all combinations of other line widths;
    \item the line red asymmetries, measured at different levels of line intensities, as tracers of the intrinsic gravitational redshift, correlates well with the corresponding line widths. The stronger correlation is seen in H$\alpha$ and H$\beta$ line widths at half and quarter maximum, whereas there is weaker or no correlation in the line-wings, indicating possible contribution of radial motions, such as inflows/outflow, to the line wing;
    \item our findings supports that the H$\alpha$ and H$\beta$ line emitting regions are having similar virialized kinematics, i.e. the gas motion is primarily driven by the gravitational force of the supermassive black hole. However, at highest velocities of the gas, which is reflecting in the line wings, there could be some signature of radial motions.
\end{enumerate}

\section*{Acknowledgements}
 Author would like to thank the anonymous referee whose comments  and suggestions helped to improve and clarify this manuscript, and to Dragana Ili\'c and Luka \v C. Popovi\'c for many discussions throughout the work on this paper, as well as for careful reading of the manuscript and suggestions for improvements. Spectral fittings presented in this publication were performed using the SUPERAST computer cluster of the University of Belgrade - Faculty of Mathematics, Department of Astronomy (http://astro.math.rs).

This publication makes use of public SDSS data. Funding for the Sloan Digital Sky Survey IV has been provided by the 
Alfred P. Sloan Foundation, the U.S. Department of Energy Office of Science, and the Participating 
Institutions. SDSS-IV acknowledges support and 
resources from the Center for High Performance Computing  at the University of Utah. The SDSS 
website is www.sdss.org.

\section*{Data Availability}

The data underlying this article are available in the article and in its online supplementary material.




\bibliographystyle{mnras}
\bibliography{example} 








\bsp	
\label{lastpage}
\end{document}